# 2D ferroelectrics and ferroelectrics with 2D: materials and device prospects

Chloe Leblanc[1,†], Seunguk Song [1,†], and Deep Jariwala[1]*

[1] *Department of Electrical and Systems Engineering, University of Pennsylvania, Philadelphia, Pennsylvania 19104, United States*

[†] These authors contributed equally: Chloe Leblanc, Seunguk Song

[*] Correspondence should be addressed. Email to: dmj@seas.upenn.edu (D.J.)

## ABSTRACT

Ferroelectric and two-dimensional materials are both heavily investigated classes of electronic materials. This is unsurprising since they both have superlative fundamental properties and high-value applications in computing, sensing etc. In this Perspective, we investigate the research topics where 2D semiconductors and ferroelectric materials both in 2D or 3D form come together. 2D semiconductors have unique attributes due to their van der Waals nature that permits their facile integration with any other electronic or optical materials. In addition, the emergence of ferroelectricity in 2D monolayers, multilayers, and artificial structures offers further advantages since traditionally ferroelectricity has been difficult to achieve in extremely thickness scaled materials. In this perspective, we elaborate on the applications of 2D materials + ferroelectricity in non-volatile memory devices highlighting their potential for in-memory computing, neuromorphic computing, optoelectronics, and spintronics. We also suggest the challenges posed by both ferroelectrics and 2D materials, including material/device preparation, and reliable characterizations to drive further investigations at the interface of these important classes of electronic materials.

## 1. INTRODUCTION

The need to address rising energy consumption, high-latency, storage limitations for data heavy computation has driven many research lines and innovations both at architecture and device levels[1-4]. These include advances in two-dimensional (2D) materials, such as group III-VI transition metal dichalcogenides (TMDs, e.g., $MoS_2$, $WSe_2$ etc.)[5-8] and mono-elemental semiconductors (e.g., phosphorene, tellurene etc.)[9-11], offering potential for energy-efficient electrical switches with ultra-scaled footprints[12] for efficient digital computing. In this context, the emergence of ferroelectricity in 2D materials and their devices, makes them more compelling for the exploration of novel electronic phenomena, high performance devices, and novel circuit architectures. Ferroelectricity in van der Waals 2D materials exists for many compounds either as distortion within the individual covalently bonded layer or in between two layers twisted at specific angles. These distortions lead to polarization of the lattice in these 2D layers along lateral (in-plane) and/or vertical (out of plane) directions allowing these materials to store charge in a non-volatile manner. This polarization can be switched via electric fields primarily which further allows for information to not only be stored but also erased and reprogrammed in a non-volatile manner.

Materials that demonstrate ferroelectricity down to atomic thicknesses are of particular interest due to their robust response, polarization adjustability, and simple structure: epitaxial perovskites, hafnia ferroelectrics, van der Waals (vdW) ferroelectrics and vdW heterostructures with a 2D crystal lattice, encompassing both monolayer and multilayer materials[13]. Understanding the mechanisms of 2D ferroelectrics and the interfacial dynamics of 2D/ferroelectric junctions is hence crucial for their incorporation in modern or future technology. For example, integrated 2D ferroelectric devices can be a compelling avenue for advancing electronic and optoelectronic systems[12, 14-16]. From non-volatile transistors to memory devices and photovoltaics, the integration of 2D semiconductors with ferroelectric materials holds immense promise in addressing pressing technological challenges while unlocking new opportunities for innovation and advancement.

Therefore, in this Perspective, we aim to elucidate the fundamental principles, recent advancements, and potential applications of 2D ferroelectric materials* and devices. We provide insights into the mechanisms governing the emergence of ferroelectricity in 2D monolayers and multilayers. Furthermore, we explore the interfacial dynamics of 2D/ferroelectric junctions and the implications of 2D materials for ferroelectric device performance and functionality. By highlighting the challenges and opportunities posed by both ferroelectrics and 2D materials, we encourage further research and innovation in this rapidly evolving field.

*Note: in this paper, “2D ferroelectric materials” represent vdW ferroelectric materials and vdW heterostructures with a 2D crystal lattice, encompassing both monolayer and multilayer materials.

## 2. 2D FERROELECTRIC MATERIALS.

Traditional ferroelectric materials such as fluorite oxides $Hf_{1-x}Zr_xO$ (HZO), perovskite oxides $PbZr_{0.2}Ti_{0.8}O_3$ (PZT), $SrBi_2Ta_2O_9$ (SBT), $BaTiO_3$ (BTO) etc. and wurtzite nitrides such as $Al_{1-x}Sc_xN$ (AlScN), exhibit spontaneous electrical polarization and non-volatile, programmable properties[17-20] via field induced switching of the polarization. Efforts to leverage these 3D ferroelectric (FE) materials in diverse fields, particularly for HZO, focus on commercially viable embedded memory devices in microelectronics. However, as electronic devices shrink in size, aggressively reducing the thickness dimensions of the 3D ferroelectric components becomes a challenge. In fact, there exists a tradeoff between thickness and optimal performance at such small sizes. This limitation arises due to chemical and structural discontinuities at the surface or interface (e.g., existence of "dead layer" or defects), as well as significant effects from screening and depolarization field, leading to less stable intrinsic polarization characteristics[20, 21] (**Figure 1a**). Moreover, achieving few-nm thick 3D FE films requires high-crystalline quality on lattice-matching substrates [22, 23] which impedes their applications in conventional microelectronics.

In contrast, 2D materials stacked through vdW bonds and having self-passivated surfaces, present a promising alternative to show in-plane (IP) or out-of-plane (OOP) ferroelectricity down to nanometer and monolayer thicknesses[24-27] (**Figure 1b-d**). Several layered 2D materials and systems can offer a platform to sustain a remnant polarization at room temperature[24, 26, 28-30] that is switchable under the field higher than its coercive fields (**Table 1** and **Figure 1d**). Furthermore, 2D FE materials exhibit a diverse range of bandgaps, spanning from metals[30-34] to semiconductors[24-27, 35-47] and insulators[28, 48, 49]. This differs from 3D FEs with bandgaps exceeding ~3.4 eV, which opens avenues for a new paradigm in 2D electronic device applications (**Figure 1c**). The discovery of a new spontaneous polarization, known as sliding ferroelectricity[32-34, 50-53], and the observation of ferroelectric metallicity[30, 31] further underscore the unique attributes of 2D ferroelectric materials. Several reports have demonstrated the viability of implementing large-scale transfer of 2D materials through low-energy processes on any desired substrates, giving rise to the possibility for the realization of diverse device structures and functionalities[54, 55]. Therefore, it is highly valuable to extensively investigate the principles governing the formation of spontaneous polarizations in 2D lattices and material design possibilities. This also encompasses the benefits of various 2D FE materials.

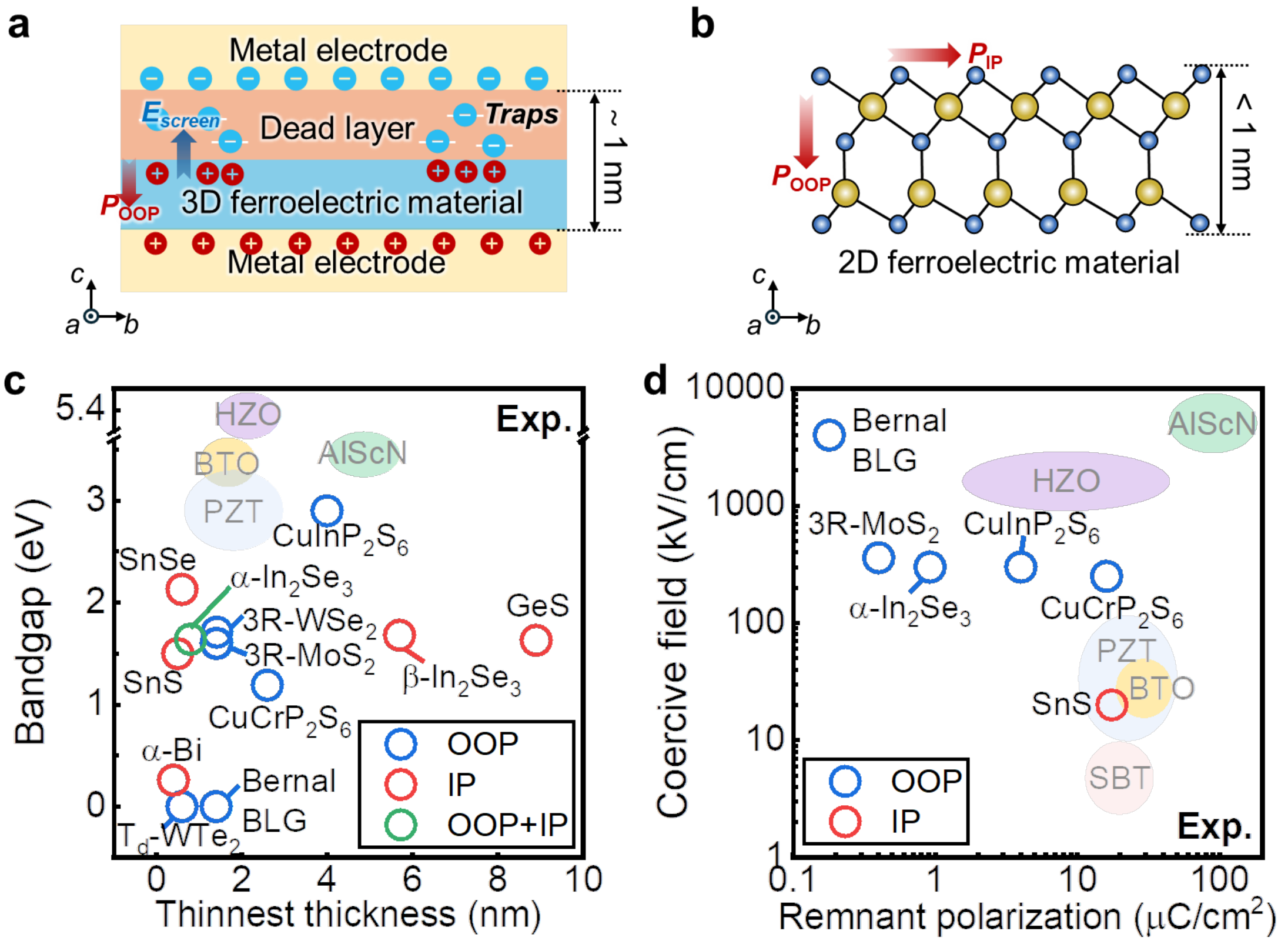

**Figure 1. 3D and 2D FE material systems.** (**a**) Schematic depicting the impact of reducing the size of a 3D ferroelectric material. A dead layer is capable of screening charges of the ferroelectric capacitance, and the resultant screening field ($E_{screen}$) can influence the out-of-plane polarization ($P_{OOP}$). The influence of this dead layer becomes more pronounced with decreasing thickness of the ferroelectric thin film. (**b**) Cross-sectional atomic model representing a 2D ferroelectric $In_2Se_3$. No dead layer is observed, and there is no discernible in-plane ($P_{IP}$) or out-of-plane polarization ($P_{OOP}$) even in monolayers thinner than 1 nm. (**c, d**) Comparative analysis of experimentally validated properties in 2D ferroelectric materials[24-37, 40, 42-45, 47, 49-53, 56, 57]; (**c**) explores bandgap and minimum thickness, while (**d**) examines coercive field and remnant polarization in 2D systems. Corresponding values for 3D ferroelectrics (PZT, SBT, BTO, HZO, and AlScN)[17-20] are presented in gray for comparison. "Bernal BLG" denotes Bernal-stacked bilayer graphene sandwiched with BN layers[32-34].

**Table 1. Representative 2D ferroelectric materials/systems and their experimentally validated properties.** Density functional theory (DFT) results are indicated in gray. Preparation methods (Prep. methods) include "ME" for mechanical exfoliation from bulk materials, "CVD" for chemical vapor deposition, and "MBE" for molecular beam epitaxy. The term "Bernal BLG" refers to Bernal-stacked bilayer graphene sandwiched with BN layers.

| Material | Minimum thickness | OOP/IP direction | Remnant or internal polarization | Coercive field or energy barrier | Band gap | Curie temp. | Prep. methods |
|---|---|---|---|---|---|---|---|
| **$CuInP_2S_6$** | 4 nm (Ref.[28]) | OOP (Ref.[28, 48]) | 4 μC/cm$^2$ (200 nm; Ref.[48]) | 300 kV/cm (Ref.[48]) | 2.9 eV (Ref.[49]) | 320 K (Ref.[28]) | ME (Ref.[48]) |
| **$CuCrP_2S_6$** | 2.6 nm (Ref.[45]) | OOP (Ref.[45]) | 16.05 μC/cm$^2$ (200 nm; Ref.[45]) | 250 kV/cm (200 nm; Ref.[45, 48]) | 1.19 eV (Ref.[47]) | 333 K (Ref.[45]) 470 K (Ref.[46]) | ME (Ref.[45, 46]) |
| **α-phase SnS** | Monolayer (Ref.[24]) | IP (Ref.[24]) | 3 μC/m (Ref.[24]) 17.5 μC/cm$^2$ (Ref. [29]) 2.62 × 10$^{-10}$ C/m (DFT; Ref. [58]) | 25 kV/cm (Ref.[24]) 20 kV/cm (Ref.[29]) 10.7 kV/cm (Ref.[59]) | 1.5 eV (Ref.[29]) | Above RT (Ref.[24, 29]) 1200 K (DFT; Ref. [58]) | CVD (Ref.[24, 29]) MBE (Ref. [59]) |
| **α-phase SnSe** | Monolayer (Ref.[25]) | IP (Ref.[25]) | 1.5 × 10$^{-10}$ C/m (DFT; Ref.[25]) 1.51 × 10$^{-10}$ C/m (DFT; Ref. [58]) | - | 2.13 eV (Ref.[25]) 0.91 eV (DFT; Ref.[60]) | 380-400 K (Ref.[25]) 326 K (DFT; Ref. [58]) | MBE (Ref.[25]) |
| **α-phase SnTe** | Monolayer (Ref. [26]) | IP (Ref.[26]) | 13-22 μC/cm$^2$ (DFT; Ref.[26]) | - | 1.6 eV (Ref.[26]) | Above RT for bilayer; 270 K for monolayer (Ref.[26]) | MBE (Ref.[26]) |
| **α-phase GeS** | 8.9 nm (Ref.[35]) | IP (Ref.[35]) | 5.06 × 10$^{-10}$ C/m (DFT; Ref.[58]) | 18.1 kV/cm (Ref.[35]) | 1.63 eV (Ref.[35]) | 6400 K (DFT; Ref. [58]) | ME (Ref.[35]) |
| **α-phase GeSe** | - | IP (Ref.[61]) | 3.67 × 10$^{-10}$ C/m (DFT; Ref.[58]) | - | 1.16 eV (DFT; Ref.[62]) | 700 K (Ref.[61]) 2300 K (DFT; Ref. [58]) | ME (Ref.[61]) |
| **α-phase Bi** | Monolayer (Ref.[36]) | IP (Ref.[36]) | 0.41 × 10$^{-10}$ C/m (DFT; Ref.[36]) | 15.7 mV/Å (DFT; Ref.[36]) | 0.26 eV (Ref.[36]) | 210 K (Ref.[36]) | MBE (Ref.[36]) |
| **α-phase $In_2Se_3$** | Monolayer (Ref. [27]) | OOP and IP (Ref.[63]) | 0.92 μC/cm$^2$ (OOP; Ref.[57]) 0.97 (OOP) and 8.0 μC/cm$^2$ (IP)(DFT; Ref.[64]) | 0.33 V/nm (OOP; Ref.[37]) 300 kV/cm (OOP;Ref.[40]) | 1.64 eV (8 nm;[56]) | 700 K (Ref.[38]) | CVD (Ref.[37-39]) ME (Ref. [27, 39, 40]) |
| **β'-phase $In_2Se_3$** | 5.7 nm (Ref.[42]) | IP (Ref.[41]) | 0.199 μC/cm$^2$ (DFT; Ref.[41]) | 0.27 eV/unit cell (DFT; Ref.[41]) | 1.68 eV (Ref.[43]) | 477 K (Ref.[41]) | ME (Ref.[41]) CVD (Ref.[42]) MBE (Ref.[43]) |
| **$T_d$-phase $WTe_2$** | Bilayer (Ref.[30, 31]) | OOP (Ref.[30, 31]) | 0.19 μC/cm$^2$ (Ref.[30]) $10^4$ e/cm (Ref.[31]) | 0.70 eV/f.u. (Ref.[30]) | 0 eV (Ref.[30]) | 350 K (Ref.[31]) | ME (Ref.[30]) |
| **1T'-$ReS_2$** | Bilayer (Ref.[44]) | OOP (Ref.[44]) | 0.07 pC/m (Ref.[44]) | 17.1 meV (Ref.[44]) | 1.6 eV | 405 K (Ref.[44]) | ME (Ref.[44]) |
| **Bernal-stacked BLG/BN** | Bilayer graphene/BN (Ref.[32-34]) | OOP (Ref.[32-34]) | 0.9-5.0 pC/m (Ref.[32]) 0.05-0.18 μC/cm$^2$ (Ref.[33]) | 0.2-0.4 V/nm (Ref.[32, 33]) | Moiré gap (Ref.[32, 33]) | 200 K (Ref.[33]), above RT (Ref. [34]) | ME (Ref.[32-34]) |
| **3R-stacked $MoS_2$** | Bilayer (Ref.[51]) | OOP (Ref.[51]) | 0.4 μC/cm$^2$ (Ref.[51]) 0.53 pC/m (Ref.[50]) | 0.036 V/nm (Ref. [51]) | 1.6 eV (Ref.[50]) | Above RT (Ref.[50, 51]) | ME (Ref.[50]) CVD (Ref.[51]) |
| **3R-stacked $WSe_2$** | Bilayer (Ref.[52]) | OOP (Ref.[52]) | 1.97 pC/m (Ref.[52]) | 0.3 V/nm (Ref.[52]) | 1.71 eV (Ref.[53]) | Above RT (Ref.[52]) | ME (Ref.[52]) CVD (Ref.[53]) |

### *2.1. Emergence of ferroelectricity in 2D monolayers.*

A non-centrosymmetric structure that separates positive and negative charge centers can induce spontaneous polarizations within a lattice. To qualify as a FE rather than a non-ferroelectric polar material, these spontaneous polarizations must be switchable along specific transition paths[13, 65]. This has been predicted and observed in various monolayers of various vdW layered 2D materials, e.g., $In_2Se_3$[27], SnSe[25], and SnS[24] (**Figure 2**). The direction of structural variation among atomic ions in the 2D monolayers determines whether OOP, IP, or both polarizations occur (**Table 1**).

For instance, as a 2D III-VI metal chalcogenide, $In_2Se_3$ exhibits distinct atomic crystal structures depending on its phases (e.g., α, β, and β’), influencing the direction and magnitude of polarization through alterations in ionic bonds. In **Figure 2a**, the α phase of $In_2Se_3$ with the atomic configuration of Se-In-Se-In-Se is illustrated[37-39, 63, 66]. It exhibits two ground polar states, determined by the motion of middle Se atomic intralayer, that feature correlated in-plane and out-of-plane polarizations. When one polarization switches, the other polarization spontaneously does so as well[63]. A theoretical calculation proposes a substantial lateral motion of Se atoms with ~100 pm[66], notably bigger than the atomic motion in conventional 3D FEs (~10 pm)[67]. In contrast to the α phase, the β phase of $In_2Se_3$ has a centrosymmetric configuration and lacks inherent ferroelectric properties (**Figure 2b**, left). However, through structural modification facilitated by the relaxation of intermediate Se atoms, the β phase can acquire in-plane polarization, referred to as the β’ phase[41-43] (**Figure 2b**, right). It is important to note that β’-phase $In_2Se_3$ can also function as an antiferroelectric material[68], where polarization is canceled out by neighboring polar domains.

Similarly, localized structural strain within a monolayer can lead to the creation of polarization, a phenomenon observed in another Group-III metal chalcogenide, GaSe[69]. Even in the typical mirror symmetric structure found in GaSe or InSe monolayers (**Figure 2c**, left), strain can introduce a local break in mirror symmetry through the sliding of Ga-Se atomic sublayers in Se-Ga-Ga-Se configurations (**Figure 2c**, right). This intralayer sliding aligns dipole moments, giving rise to correlated OOP and IP polarizations within the monolayer[69].

A-phase group-IV metal chalcogenides , exemplified by[24, 25, 29, 35, 59, 61] , are established as 2D FE with inherent IP polarization (**Figure 2d**) that arises from the distorted puckered structure formed in the armchair orientation of a highly anisotropic crystalline structure. Extensive experimental investigations have focused on the spontaneous electric polarizations and lattice strains within the realm of < 15 layers,[24, 29, 59]. Their potential as 2D multiferroic materials, combining ferroelectric and ferromagnetic orders, has also been suggested[70]. Monolayer α-phase Bismuth (Bi) also shows a crystal structure similar to SnS (**Figure 2d)**[36]. Within the 2D Bi crystal, there is a simultaneous occurrence of ordered charge transfer and regular atom distortion between sublattices. Additionally, the weak and anisotropic *sp* orbital hybridization between Bi atoms results in a buckled structure with broken inversion symmetry, leading to IP electric polarization along the armchair orientation.

**Figure 2e** displays the atomic schematic of vdW FE layered thiophosphate of $M^{1+}M^{3+}[P^2S^6]^{4-}$, where $M^{1+}$ = Ag, Cu, and $M^{3+}$ = Cr, In. Especially, $CuInP_2S_6$ (Refs.[28, 48]) and $CuCrP_2P_6$ (Refs.[45, 46]) are representative FE materials where the Cu and In (or Cr) atoms displace in opposite directions within an S framework with octahedral voids. Changes in symmetry, such as the order of displacement from the center of the Cu lattice and the displacement of cations in the center symmetry of the In lattice, lead to spontaneous polarization along the OOP direction in the FE phase. The bandgap of $CuInP_2S_6$ (~2.9 eV)[49] is considerably larger than most 2D FE materials (**Table 1**). This distinctive feature sets it apart among 2D materials, making it a rare candidate capable of serving as a ferroelectric insulator in 2D channel-based devices, especially in ferroelectric field-effect transistors (FeFETs)[71, 72].

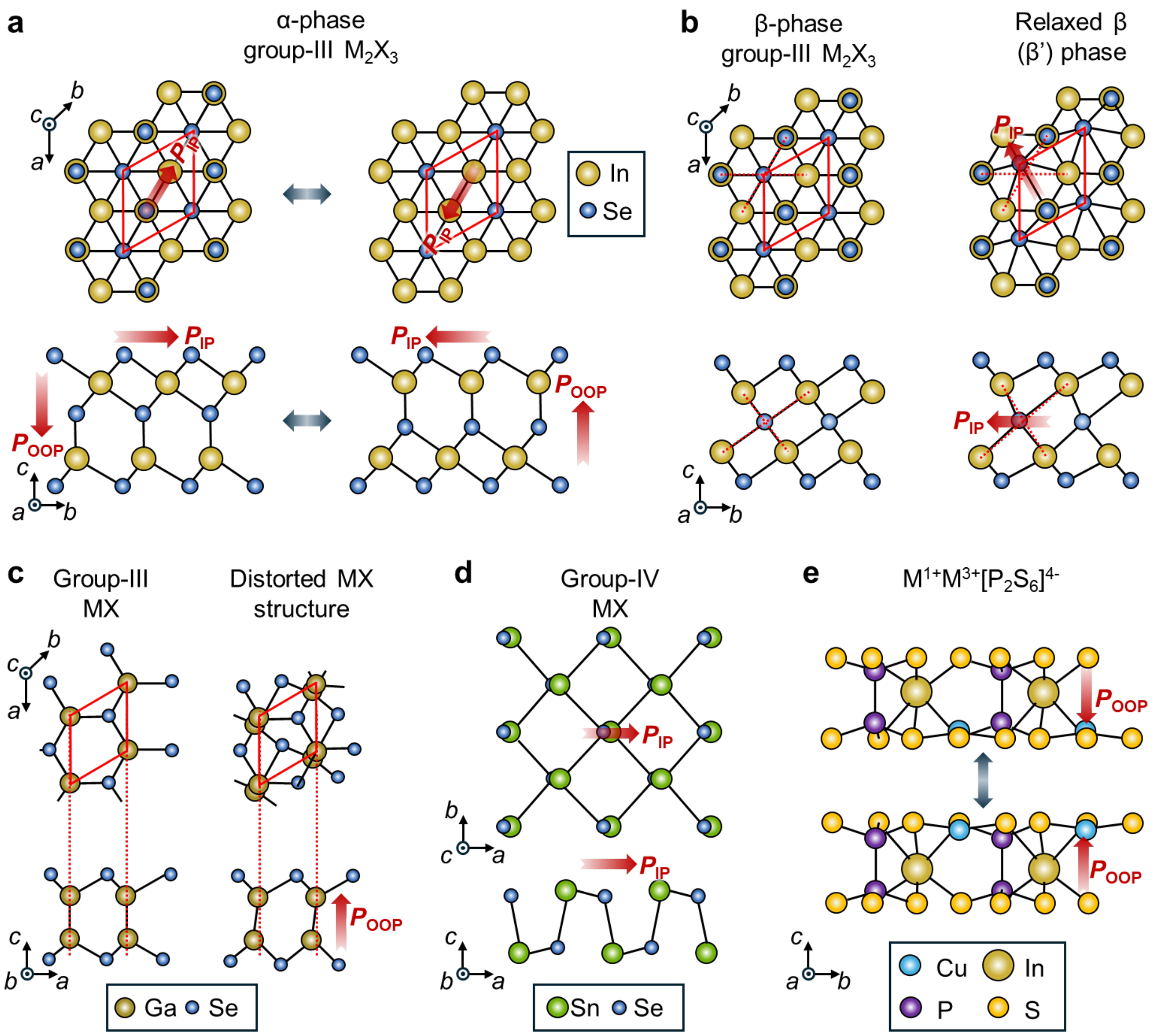

**Figure 2. Spontaneous polarizations in 2D (anti)ferroelectric monolayers.** (**a**) Atomic configuration of the α phase of $In_2Se_3$. The ferroelectric α-phase $In_2Se_3$ exhibits both in-plane and out-of-plane polarization, correlated and switchable upon each other. (**b**) Left: Centrosymmetric configuration of the β phase of $In_2Se_3$. Right: The β' phase, achieved through structural modification, showing in-plane polarization. (**c**) Left: Mirror symmetric structure of GaSe or InSe monolayers. Right: Localized structural strain introduces a break in mirror symmetry through sliding of Ga-Se atomic sublayers, leading to correlated out-of-plane and in-plane polarizations. (**d**) Crystal structure of group-IV metal chalcogenides (SnS, SnSe, GeS, GeSe) in the α phase, exhibiting inherent in-plane polarization. The armchair orientation within the anisotropic crystalline structure gives rise to unique polarization. (**e**) Atomic schematic of vdW ferroelectric layered thiophosphate $M^{1+}M^{3+}[P_2S_6]^{4-}$ ($M^{1+}$ = Ag, Cu, $M^{3+}$ = Cr, In).

### *2.2. Ferroelectricity in 2D multilayers and artificial 2D ferroelectricity.*

Robust atomic-scale ferroelectricity n stacked 2D multilayers can be induced by charge redistribution through the hybridization of occupied and unoccupied states between the multilayers or the net charge transfer across the vdW interface can lead to the generation of OOP polarization. In particular, the structural transformation into a mirror image, accompanied by the reversal of charge redistribution, can occur with relative shifts between 2D monolayers.

This "sliding ferroelectricity" has been observed in bilayers of h-BN[73] (**Figure 3a**) and TMDs (e.g., $MoS_2$, $WSe_2$)[50, 51, 53] in rhombohedral-stacked (3R) structures (**Figure 3b**). In an ideal 3R bilayer, the two layers are stacked parallel with no misalignment of the lattice (AA stacking). As these two layers slide against each other along the vdW interface, different stacking orders can occur, creating what is known as 3R polytypes in crystallography. Specifically, in the AB and BA stacking order, there is a possibility of charge transfer leading to the formation of an OOP polarization. Similarly, for $WTe_2$ or $ReS_2$, the charge redistribution between the bilayers is induced by sliding in-plane translation of the distorted 1T structure (1T' or $T_d$ phase)[31, 44].

The switchable polarization through 2D crystal symmetry translation also occurs at multi-stacked vdW interfaces within 2D multilayers beyond the bilayer configurations[50, 74]. It has been demonstrated that a switchable polarization of ~0.5 pC/m can sustain at each 2D TMD multilayer interface, enabling the support of high charge densities up to $\sim 10^{13}$ $cm^{-2}$ (Ref.[50]). OOP polarization and IP conductivity can coexist due to the localized polarization states and charge carrier depolarization on both layers. Hence, the number of layers in a vdW 2D material, along with the coupling of dipoles between each layer, not only governs sliding ferroelectricity but also presents diverse possibilities for novel multiferroic materials combined with different vdW layers or other other ferroic orders (e.g., ferromagnetism).

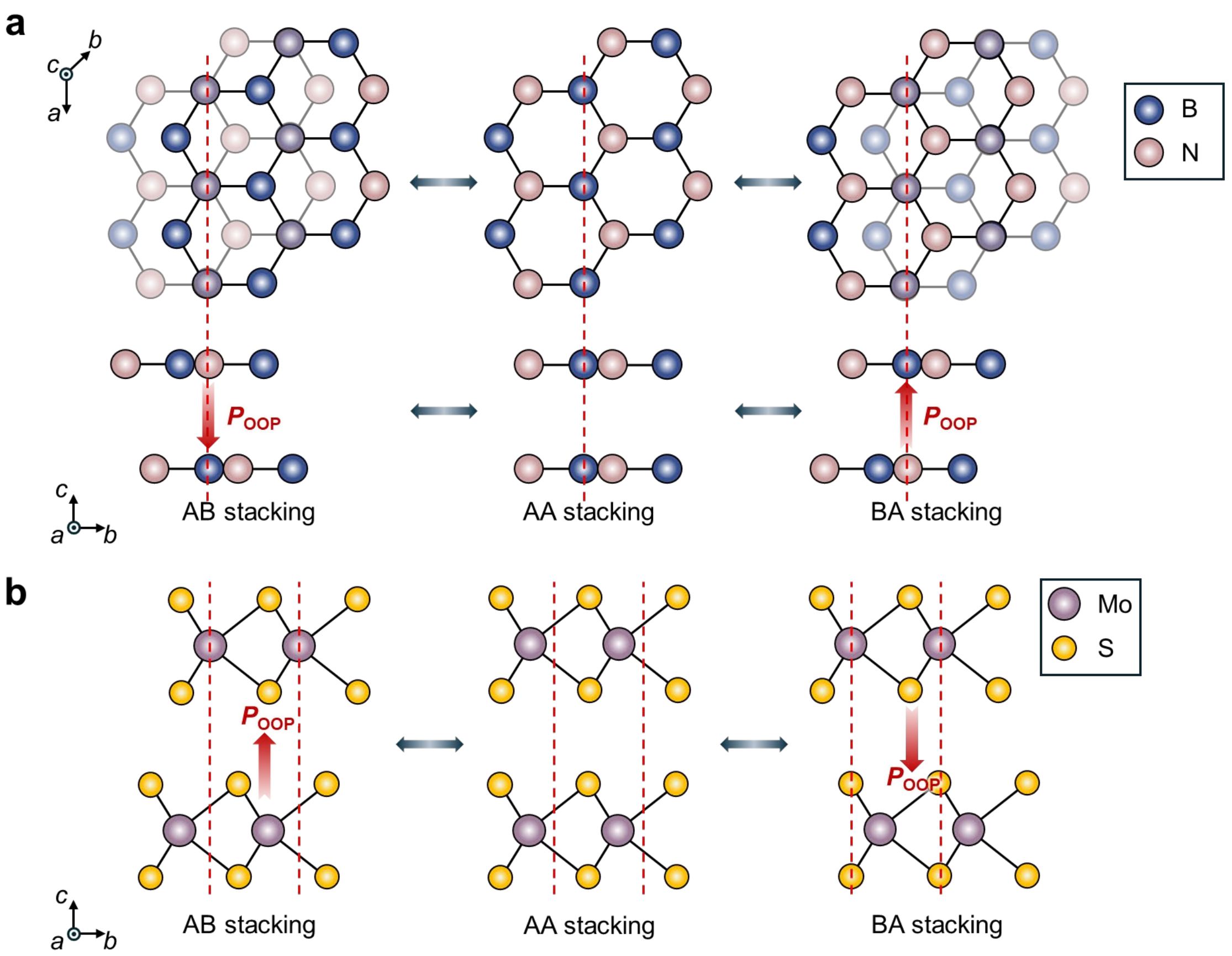

**Figure 3. Structural formation in stacked 2D multilayers and their polarizations.** (**a**) Ferroelectric switching in bilayers of h-BN. (**b**) Ferroelectric switching in 3R-structured transition metal dichalcogenides (e.g., $WSe_2$, $MoSe_2$, $WS_2$, and $MoS_2$) . The relative shifts between layers lead to the generation of OOP electric dipoles, accompanied by mirror image transformation and reversal of charge redistribution.

According to the working principle of sliding ferroelectricity, it is possible to artificially produce and control ferroelectricity using non-FE 2D monolayers (**Figure 4**). For example, strategies to induce sliding ferroelectricity by substitutional doping or stacking different 2D TMD layers have been proposed. In the case of InSe, the piezoelectric constant can be increased by one order of magnitude higher after doping Y (~7.5 pm/V), and the IP and OOP polarizations achieved by eliminating stacking defects due to interlayer compression and continuous interlayer pre-sliding, and by subtle distortion in rhombohedral configuration[75] (**Figure 4a**). Additionally, sliding ferroelectricity can be achieved in a 2H structure rather than a 3R by stacking two different single-layer TMDs. In a vertical heterostructure of $MoS_2/WS_2$ bilayer, the different transition metals of Mo and W break the symmetry transformations that exchange atoms in the bilayer as well as the inversion center, resulting in OOP polarization for both structures[76] (**Figure 4b**).

Sliding ferroelectricity can also be induced in twisted bilayer structures. **Figure 4c** depicts the schematic representation of a twisted bilayer's structure, formed by rotating two 2D monolayers with hexagonal lattices through a small angle. In this schematic, regions where the lattices of the two bilayers are nearly aligned and exhibit inversion symmetry form an AA stack. These aligned structures create an in-plane Moiré pattern, appearing relatively bright. Surrounding these AA regions are AB or BA regions with broken inversion symmetry, each representing a ferroelectric domain intersecting the OOP polarization. Consequently, as the domain walls move, charge transfer occurs at the interfaces between the layers, leading to the exhibition of OOP ferroelectricity in response to the switching of these polarizations[33]. The movement of the domain wall has been observed through back-scattered channeling contrast electron microscopy in two twisted bilayers of $MoS_2$ sandwiched between h-BN layers[77]. The imaging reveals variations of domain sizes within the $MoS_2$, and by extension the dependence of domain behavior on lateral dimensions. Furthermore, the motion of these domain walls and alterations in electrical polarization are corroborated by the OOP electric field[77].

Another example using the Moiré potential is found in the Bernal bilayer graphene sandwiched by the h-BN layers[32-34] (**Figure 4d**). The precise controlling of the angle between the Bernal bilayer graphene and beneath h-BNs gives rise to the Moiré potential with localized ferroelectric domains. Hence, the spontaneous ferroelectric polarization in this structure ranges from ~0.9 to 5.0 pC/m (Ref.[32]) or ~0.05 to 0.18 $\mu C/cm^2$ (Ref. [33]), which is notably large for a 2D system (**Table 1** and **Figure 1d**). The electrons can fill in the asymmetric and localized Moiré potentials by the response to the displacement field, allowing for the unconventional hysteresis of the electrical conductance (i.e., electronic ratchet states)[34]. Therefore, the expanding family of Moiré 2D heterostructures, formed by twisting 2D nanosheets with diverse morphologies, is expected to offer further non-volatile electrical properties that can be beneficial for a range of electrical device architectures[34].

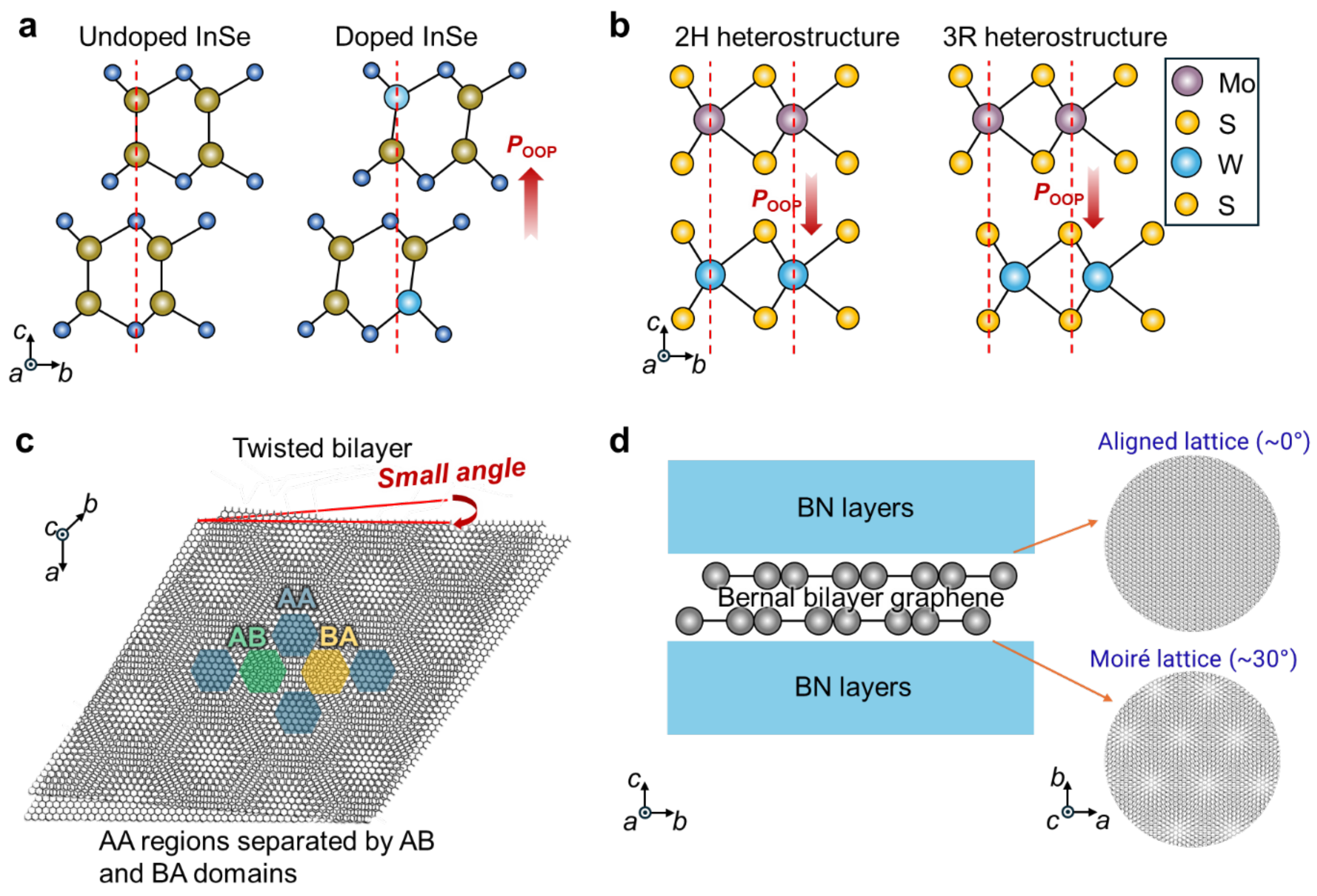

**Figure 4. Engineering of sliding ferroelectricity in 2D heterostructures.** (**a**) InSe sliding ferroelectricity enhanced by Y-doping. (**b**) Vertical $MoS_2/WS_2$ bilayer ferroelectricity. Sliding ferroelectricity in both 3R and 2H structures emerge due to broken inversion symmetry induced by different transition metals (Mo and W). (**c**) Schematic of twisted bilayer structure illustrating formation domain of AB and BA stacked regions, leading to OOP ferroelectricity. (**d**) Bernal bilayer graphene between h-BN layers exhibiting Moiré potential-induced localized ferroelectric domains.

## 3. DEVICES.

As discussed in **Section 2**, the numerous 2D materials and systems that exhibit ferroelectric properties possess switchable polarizations, leading to their possible implementation in various electronic devices for logic and memory applications[78]. Their vdW interfaces allow fabrication of highly scaled and densely packed atomically thin nanodevices. Moreover, novel heterostructures made of stacked vdW materials display new functionalities and modulable properties[79]. Various two- and three-terminal ferroelectric device structures have been extensively investigated in research settings (**Figure 5**). Furthermore, different 2D devices for logic, memory, and sensing applications can be stacked together to produce monolithically integrated 3D systems. Ferroelectric devices based on 2D materials draw particular attention for their potential to address the bottleneck of concurrent computation paradigm. For instance, traditional von Neumann computing architectures process tasks and store memory in separate units, which implies that the exchange of data between the two locations consumes both energy and time. Depending on the amount of displaced data, memory access may even expend more energy than the task at hand[79]. For instance, accessing an off-chip DRAM consumes nearly thousand times more power than a 32-bit floating point operation[80]. Today, the drastically increasing demand in computing power has made this collateral latency intolerable—despite enhanced individual circuit component performances with every evolving generation of devices. Ferroelectric non-volatile memory (NVM) technology could be the answer to this challenge as it enables both low-power memory and logic units, which could lead to embedded memory and in-memory computing applications that are vastly energy efficient over traditional von Neumann architectures[17, 81].

However, the current challenge lies in the utilization of the 2D ferroelectric materials through large-area, wafer-scale synthesis, and integration methods (as outlined in **Section 6**). Given that some of the 3D FE (e.g., $Hf_xZr_yO$) are commercially available, the large-area FE device applications employing 2D materials are anticipated to manifest with the use of 3D FEs with non-ferroelectric 2D semiconductors. Furthermore, the stacking of a 2D non-FE material semiconductor with a 3D FE material introduces a range of possibilities in terms of device applications. 2D FE systems inherently possess a coercive force and a remnant polarization more than an order of magnitude lower than those observed in 3D FE materials such as HZO or AlScN, as illustrated in **Figure 1d**. Simple calculation indicates that the magnitude of the current density $n \approx P_r/q$ (where $P_r$ is the remnant polarization and $q = 1.6 \times 10^{-19}$ C is the elementary charge) varies by a factor of 10 or more for 3D FE. The substantial coercive force ($E_c$) also holds an advantage, providing a sizable memory window MW $\approx E_c \times$ thickness of FE for FE field effect transistor (FeFET).

This review therefore aims to comprehensively examine available FE devices with 2D materials. This section reviews the different device configurations, followed by the advantages associated with the use of 2D materials in ferroelectric devices. For two-terminal and three-terminal FE NVM devices, the performances of 2D FE devices are also compared with those of the 3D FE devices (**Figures 6, 7**). Note that, the NVM devices are assessed with specific key performance metrics: how long a memory state can be retained (retention time), the number of cycles before the binary states decay to indistinguishable (program/erase endurance), the drain current between the program and erase modes (ON/OFF ratio), the potential for integration (scalability), the voltage separation between the ON and OFF switching voltages (memory window), and the added capacitance values across the device (C-V characteristics). These criteria are not optimized in current commercial memory systems, such as Flash and dynamic random-access memory (DRAM), which tend to require large supply voltages and power or suffer from low write endurance and short retention[82].

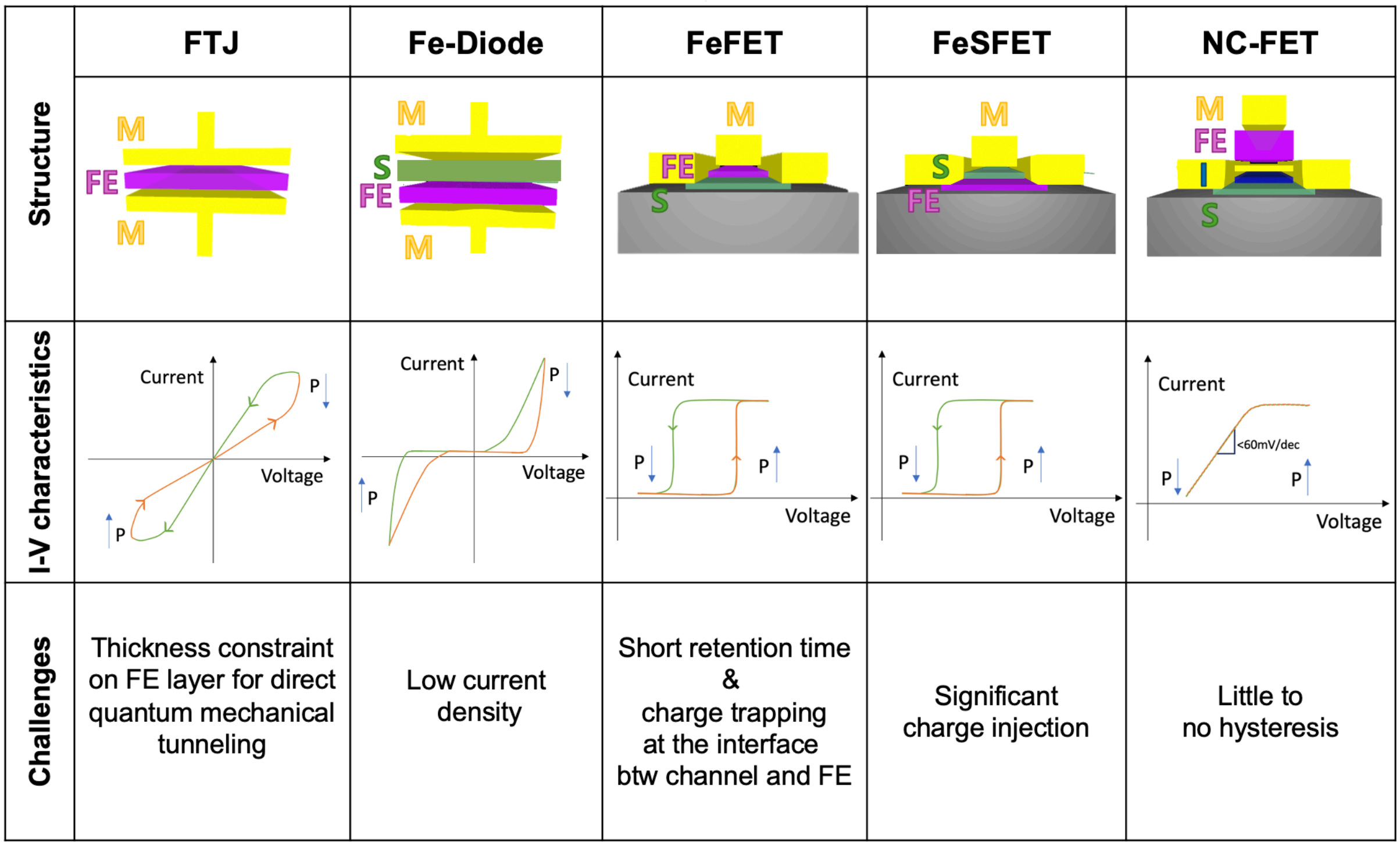

**Figure 5. Structure and properties of common 2D FE devices.** Abbreviations in the *Structure* list device layers from top to bottom (M: metal, F: ferroelectric, S: semiconductor, I: insulator). Ferroelectric field-effect transistor (FeFET), ferroelectric tunnel junction (FTJ), ferroelectric diode (Fe-Diode), ferroelectric semiconductor field-effect transistor (FeFET), and negative capacitance field-effect-transistor (NC-FET) devices are shown. The ferroelectric materials, insulators, and metals can be 2D materials. In the Current-Voltage (*I-V*) characteristics, the orange line represents a forward voltage sweep, and green line represents a backwards sweep. $P\uparrow$ and $P\downarrow$ symbols stand for polarization up and polarization down respectively.

### *3.1. Two-terminal devices.*

In two-terminal configurations, contacts modulate both the input signal that switches the ferroelectric polarity and the output signal. The polarization status of the film then determines the electron tunnelling transmittance and resistive switching[83]. This allows the devices to be smaller and by extension more attractive for highly dense integration and low power consumption. They also have the advantage of easier manufacturing over more complex multiterminal devices—even though the later have more flexible operations. Common examples of such ferroelectric data storage cells are tunnel junctions (FTJ) and diodes (Fe-diodes) (**Figure 5**). FTJs consist in a thin FE layer sandwiched between two metal electrodes. The application of an electric field to the cell switches the FE film polarization, which changes the barrier height that electrons must overcome to tunnel across the device[84]. This tunneling electroresistance in FTJs can be enhanced by notably adding a dielectric layer to the device[85].

Fe-diodes are similar to FTJs, with the important distinction that the single ferroelectric layer is substituted for both a semiconductor and a ferroelectric layer. This adds a layer of difficulty in the design, because the band alignment at the semiconductor/ferroelectric heterojunctions interface now determines the conductance through the device. The polarization orientation of their FE layers modulates the Schottky barrier at the interface, which tunes the resistivity of the device[44]. They exhibit and non-linear current-voltage characteristic, and their rapid switching speed between polarization states allows for fast read and write operations in memory devices. Decreasing cell dimensions is particularly appealing for technological applications. Devices with nanometer-thick FE layers would allow for higher efficiencies and storage densities. However, the miniaturization of these devices causes degradations in their performance due to local defects, strain, and the formation of 'dead layers'[86].For instance, the assembly of a Fe-diode junction in ambient conditions allows for the presence of unwanted water molecules at the interface, which at those dimensions lead to a depolarization effects[87]. In FE capacitors, devices are also subject to destructive readout process, where sensing disrupts the stored data and write-back is required.

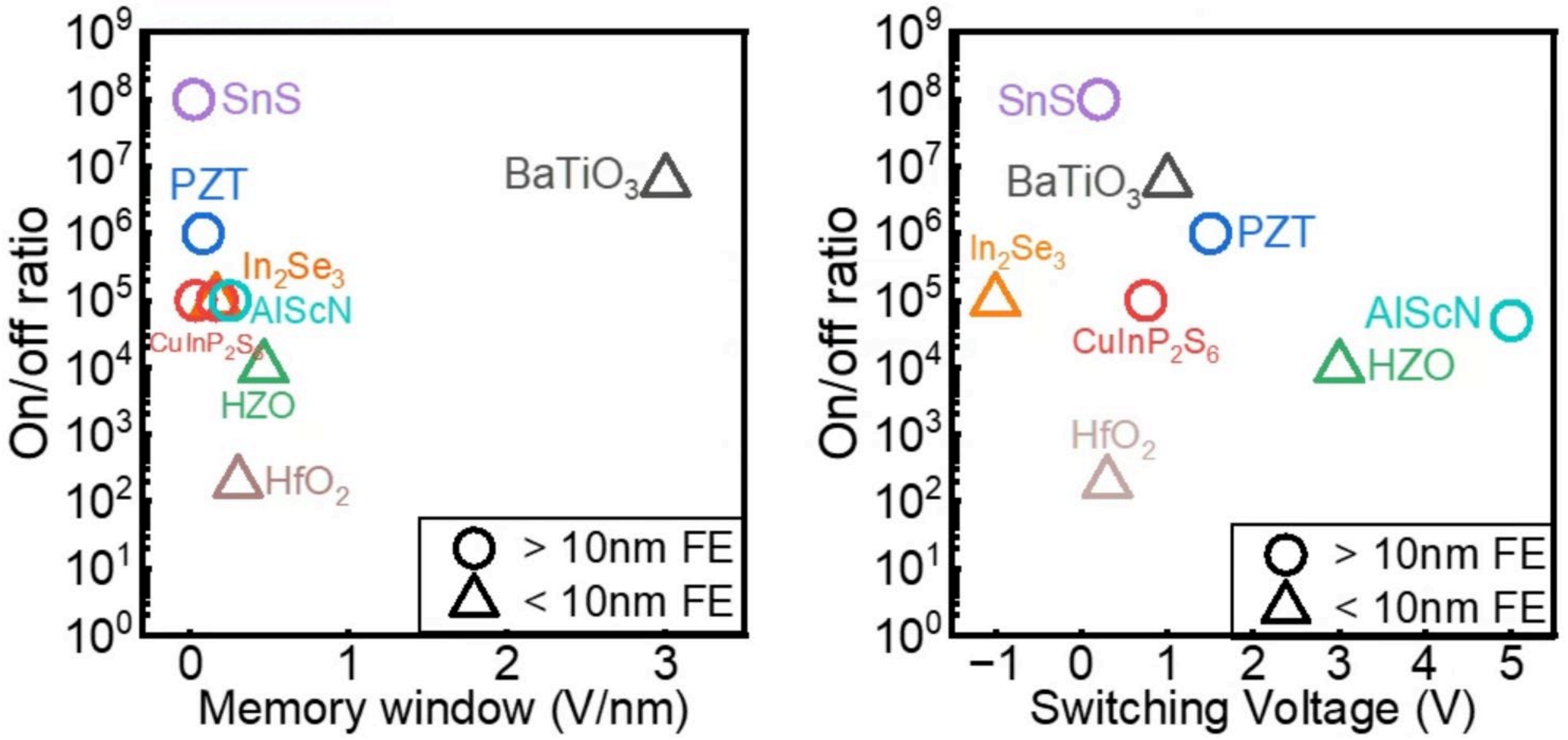

**Figure 6. Reliability plots of reported Fe-diode and FTJ performances**. High ON/OFF ratios, large memory windows and low switching voltages are preferred (Ref.[88-100]). Triangular symbols indicate devices with FE layers < 10nm, and circular symbols indicate devices with FE layers > 10nm.

### *3.2. Three-terminal devices.*

Three-terminal devices separate themselves from two-terminals based on readout process: their input and output signals are allocated to separate terminals, which enables non-destructive reading and lower power consumption[101]. We here briefly discuss the mechanisms of ferroelectric field-effect transistors (FeFET), ferroelectric semiconductor field-effect transistors (FeSFET) and negative-capacitance field-effect transistors (NC-FET).

FeFETs as a device concept turn 50 years old in 2024[102]. Their architecture combines the non-volatility of stored charge in ferroelectrics with the gain of a transistor by substituting the dielectric layer for a ferroelectric material that controls the channel conductance[103]. In its ON state, the ferroelectric layer introduces electrons in the adjacent channel to increase conductance. To revert to an OFF state, the ferroelectric layer's domains must switch in the opposite polarity, which is performed by the application of a negative voltage below the threshold at the gate in the case of an n-type FeFET or a positive voltage above the threshold in the case of a p-type FeFET[104]. In 3D FeFETs, the voltage required to switch the device is often high due to the ferroelectric layer being too thick. This further motivates the effort to achieve 2D FeFETs, in which the FE layer dimensions have been decreased to the nanometer scale[87]. Multidomain polarization states in some ferroelectric materials even allow for several logic states and precise tuning of the drain current, hinting to the possibility of analog devices[105]. Two major reliability concerns that have held back FeFETs from widespread commercialization are 1) insufficient endurance, caused by gate leakage current, leading to the ON and OFF states convergence, and 2) insufficient retention due to large depolarization fields related to depletion region of semiconductor and charge trapping at the ferroelectric/semiconductor interface[104,106]. These affect the information that can be stored and rewritten in the FeFET. The intrinsic properties of the 2D FE layer directly correlate with the scalability and lifetime of the device. To minimize leakage, 2D layers with low capacitance are preferable. A low permittivity translates to fast switching speeds, whereas a low $E_c$ decreases charge trapping during writing[87].

The FeSFET with 2D FE semiconducting channel can address the above-mentioned challenges of FeFET, specifically insufficient retention and large depolarization field. Its structure is comparable to the FeFET, with the substitute of the ferroelectric material becoming the channel, and non-FE dielectric induces field effect on channel conductance. The polarized charges then accumulate at both the top and bottom of the semiconductor[107]. FeSFETs have been demonstrated with 2D semiconducting FE materials such as $\alpha$-$In_2Se_3$, InSe, and CIPS[108, [109]. Record FeSFET performances include $10^8$ ON/OFF ratio, $10^7$ cycles endurance, and $10^{-2}$ A/µm$^2$ current density (**Figure 7**). Some devices have even merged a 2D FE channel with a FE dielectric, such as InSe/CIPS[71].

NC-FETs are also structurally similar to FeFETs (**Figure 5**), but the negative capacitance of their ferroelectric insulators results in a large voltage amplification at the oxide/semiconductor interface. The capacitance is stabilized in a single state, which translates to the memory window of the device shrinking to near null and subthreshold swing (SS) decreasing below Boltzmann's 60 mV/dec limit[110, 111]. These devices are more auspicious to 2D channels than 3D, as 2D materials have more stable capacitances[112]. For instance, AlScN-based NC-FETs with 2D $MoS_2$ channels are able to reach SS values down to 30.7 mV/dec[113]. P($VDF_{0.75}$-$TrFE_{0.25}$), HZO, and CIPS NC-FETs with the same channel were demonstrated with 11.7 mV/dec, 6 mV/dec, and 28 mV/dec SS respectively[112, 114, 115]. NC-FinFETs have also been reported with extremely low SS, notably with $BiFeO_3$ and HZO as FE layers[116, 117]. However, the application of NC-FET is not NVM, but low-power logic transistor, so there should be the minimization of hysteresis (i.e., MW) and maximization of switching speed. To pursue the negligible MW, non-FE dielectric layer should be inserted within gate stack to stabilize the NC effect[113, 118].

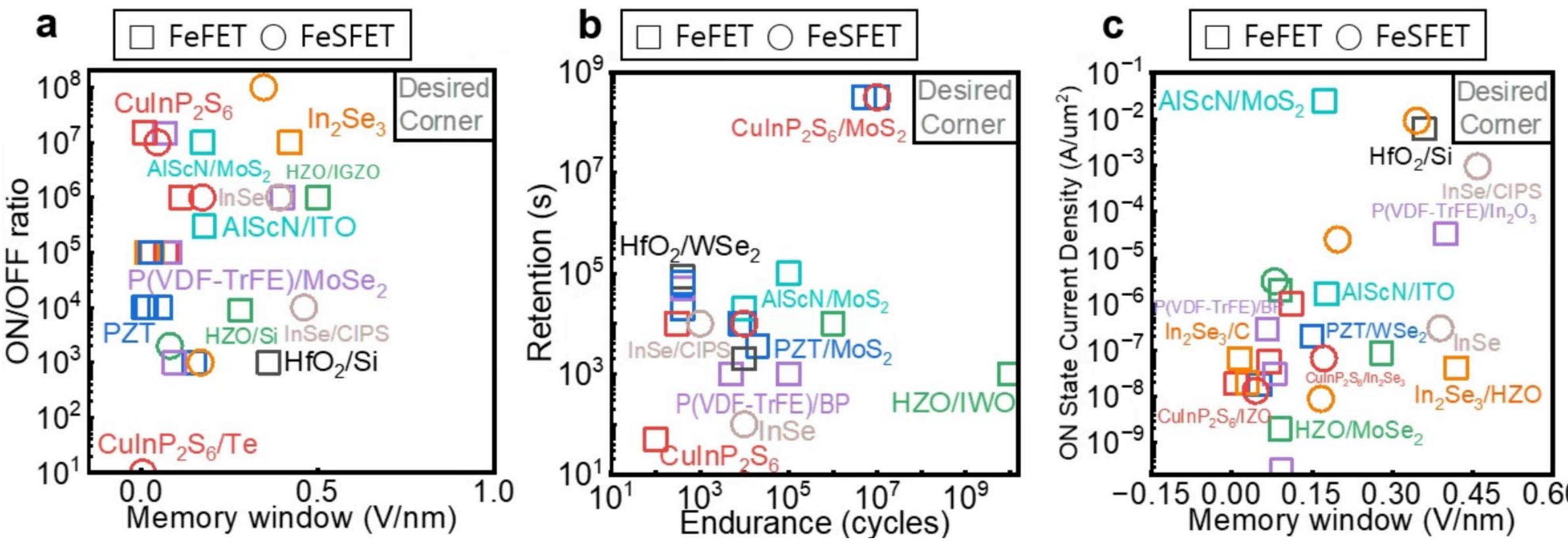

**Figure 7. Performance and reliability plots of reported FeFETs and FeSFETs** (Ref.[7, 71, 72, 79, 106, 108, 119-148]). FeFETs are square symbols and FeSFETs are circular symbols. High performing devices aim to have a large ON/OFF ratio, memory window, retention, and endurance over time.

### *3.3. Benefits of FE devices with 2D materials.*

Devices made of 2D non-FE semiconductor channels and 3D FE exhibit enhanced memory non-volatility over those with 3D non-FE channels (such as Si-based FeFETs). This superiority arises from the reduced depolarization field in FeFETs with 2D channels. The depolarization field ($E_{dp}$) in these FeFETs is governed by the equation[149, 150]

$$E_{dp} = \frac{-P}{\varepsilon\,(\frac{C_{IS}}{C_{Fe}} + 1)} \tag{1}$$

, where $P$ represents the polarization in the FE material, $\varepsilon$ is the permittivity of the FE material, $C_{Fe}$ is the capacitance of the FE, $C_{IS}$ is the effective capacitance of the semiconductor channel. As is the case for ultrathin channels (i.e., 2D materials), the depolarization field ($E_{dp}$) converges to zero when $C_{IS}$ approaches infinity. Another factor contributing to retention loss is charge trapping at the interface of FE and non-FE materials. The expected retention time ($t$) for a FeFET device with remnant polarization ($P_r$), leakage current ($I_{leak}$), and trapping probability ($\alpha$) can be estimated as[150];

$$t = \frac{P_r}{\alpha I_{leak}} \tag{2}$$

Si-based FeFETs devices with insulating buffer layers (e.g., $SiO_x$) may exhibit a high $\alpha$ due to the buffer layer causing local charge trapping, resulting in a shorter retention time[149, 150]. In contrast, the use of 2D channel in heterostructures may allow for the maintenance of a pristine vdW interface without an oxide buffer layer, which leads to a longer retention time in 2D channel FeFETs due to the low trapping probability. Therefore, utilizing a 3D FE layer with a 2D vdW channel can be an effective combination for high-retention memory devices. This takes advantage of the atomically thin nature of 2D layers and the high $P_r$ in 3D FE materials (see **Figure 1c, d** for their comparisons).

Furthermore, the advantages of 2D FE materials over their bulk counterpart have widened the field of ferroelectric-semiconductor junction possibilities, which leads to novel 2D/2D structures such as FeSFETs or $Fe^2$-FETs with a double-stacked vdW FE materials (e.g., $In_2Se_3/CuInP_2S_6$ structure)[88, 106]. These 2D devices with 2D FE materials can be useful for device miniaturization because FE material itself can be atomically thin. Moreover, 2D $Fe^2$-FETs can have overall enhanced polarization due to dipole attraction between two FE materials, leading to a wider hysteresis window and longer polarization[151].

## 4. TECHNOLOGICAL APPLICATIONS

Ferroelectric devices are applied in a wide variety of current and emerging fields, a few of which we highlight below. Their operation relies on using the malleable electric polarization of ferroelectric materials to store information in devices that can then perform tasks such as logic operations. We focus on memory and neuromorphic computing, as well as optoelectronics and spintronics[152].

### *4.1 In-memory computing (logic circuits, etc.)*

FE materials can improve the performance in existing memories. For instance, for their combined memory and logic (switching and amplifying) functions, FeFETs inside Flash memory could provide a higher endurance over time[153] as well as lower switching energy. In circuit design, versatile FeFETs with varying polarities serve as non-volatile logic gates[72]. Recently, FeFETs with different polarities have been realized using 2D semiconductor channels such as $MoS_2$ and $WSe_2$ in conjunction with 3D ferroelectric AlScN gates[15, 16]. By considering the energy alignment between the 2D channel and the contact electrodes, the FeFETs devices can exhibit *n*-type, *p*-type, and ambipolar transport characteristics[16].

FE polarization retention is also used in backup/restore modules of flip-flop schemes, where the system's state can be stored[154]. Similarly, FTJs are a convenient storage element in 1T/1C random-access memory structures (RAM) for commercial applications (**Figure 5**)[155,103]. Here again, FeFET flip-flops are more efficient than 2 terminal FE-capacitor flip-flops, as the backup-restore module is simplified and faster[156].

Emerging technologies such as autonomous systems also benefit from 2- and 3-terminal FE devices. These systems that harvest energy from the environment without the need of a battery charge are of particular interest in countless research fields, ranging from autonomous cars to surgical robots. However, they require constant state backups in the event of a power failure, which ferroelectric platforms can ensure[154].

The intrinsic switching stochasticity in FE memory units even finds uses in random number generators, which are key to cryptography and statistical systems[155].

### *4.2 Neuromorphic computing*

The human brain contains $\sim 10^{11}$ neurons that communicate in parallel through electrical signals across $\sim 10^{15}$ synapses[157]. Synapses' low power consumption (1–10 fJ per event) has inspired neuromorphic computing, which can be extended to a ferroelectric behavior[155]. For instance, a given synapse's coupling strength, or weight, between adjacent neurons is modulated by electrical firing frequency and patterns. Likewise, the polarization of FE domains in an FTJ is altered by voltage pulses. A change in amplitude or frequency results in a different percentage of switched domains, hence a different resistive state[158]. A parallel can therefore be drawn between the resistivity of the FTJ and the weight of a synapse. Then arranging a crossbar array of FTJs constructs a hardware network where each cell has its own weight[103]. 2D Fe-diodes, memristors, FeFETs, and FeRAMs have also been described as neuromorphic[152]. For example, the programing of gradual states in some FeFETs can be used to perform analog vector-matrix multiplications in artificial networks[81, 103, 159].

Recently, Moiré heterostructures have also entered the field of neuromorphic computing. Instead of relying on ion migration or atomic displacement, Moiré synaptic transistors (MST) exploit the ratcheting states to tune conductance in their heterostructures (Section 2.1.2). Dual-gated bilayer-graphene/hBN (BLG/hBN) Moiré heterostructures in particular have demonstrated unique gate-tunability, high endurance and low power consumption[34]. These also achieve bidirectional synaptic threshold sliding controlled by voltage, which points to a facile integration in hardware.

### *4.3* ***Optoelectronics (light-sensitive synapses, etc.)***

Optoelectronic memory devices on the market today are composed of physically separate sensing and memory units. Owing to their atomically-thin nature, and strong photon–matter interaction, 2D TMD semiconductors are extensively explored in photoelectric detection. A local FE polarization can enhance their collection efficiency by adjusting the band structure at the interface and suppressing recombination of photogenerated carriers[12]. This concept can be applied to novel NVM devices: the photomemristor is a 2-terminal structure whose responsivity is configured by photoexcitation and stored in the device. Its behavior displays a hysteresis pattern with an ON/OFF switching ratio of $10^3$ following high-intensity light exposure. By doing so, photomemristors use photoresponse as a state variable and merge light-sensing, logic and memory in the same unit, thereby simplifying processing in optical circuits[160]. Polarization-dependent photoresponse was found in other heterostructures such as 2D BP/PZT and 2D $WSe_2$/h-BN photonic memristors, the latter which has been reported as having 128 distinct photoresponsive states and an ON/OFF ratio of $10^6$ [14,161]. Light-modulated polarization in $BaTiO_3$ has also been applied to optoelectronic FeFETs with monolayer $MoS_2$ channels to achieve wavelength-dependent neuromorphic vision sensor[162]. In 2020, an optoelectronic FTJ was even demonstrated using α-In2Se3, wherein light information was store in the ferroelectric's domain walls. The resulting device displayed an ON/OFF ratio of 104 and broadband storage covering wavelengths above 900 nm[163]. Further work in FE optoelectronics should focus on back-gated devices and transparent films to avoid FE absorption of incident light.

### *4.4 Spintronics*

Spintronic-based memories revolve around ferromagnetic materials such as CoFeB and MgO, in which the surface magnetic moment can be tuned by an electric field[164,165]. The direction of magnetization then dictates ferromagnetic material behavior, much like the electric dipole in ferroelectrics. An example of spintronics technology is the magnetic-tunnel-junction (MTJ) structure, used in MRAM and magnetic sensors[165]. Its design consists in a thin tunnel barrier film sandwiched between two ferromagnetic layers. The magnetic moment can be flipped upward or downward in the top layer, whereas it is kept fixed in the lower layer. The resistance of the MTJ varies according to whether the magnetization direction for the recording layer is parallel to that for the reference layer or not. 2D materials-based functional layers could improve thermal stability and control over MTJ tunneling barriers. So far, 2D graphene based MTJs have displayed a tunneling magnetoresistance ratio of 31%, but an experimental study suggests this could be as high as 100%[161]. By adding a ferroelectric barrier to an MTJ, the spin polarization becomes dependent on both tunneling magnetoresistance and tunneling

electroresistance—making the junction a four-state resistance device[166]. Spintronics present a wide range of applications, such as mass-data storage, sensors, quantum computation, and electron entanglement in semiconductor devices.

Multiferroic materials, which display more than one ferroic order (ferroelectricity, ferromagnetism or ferroelastic) open new possibilities to modulate material behavior in multi-functional, multi-terminal devices. In the cases where the ferroelectric and ferromagnetic orders are coupled, magnetic moment can control FE properties and vice versa[167]. Such multiferroic heterostructures are rare, especially at the 2D scale[70, 168]. Nonetheless, $CuCrP_2S_6$ seems a promising 2D multiferroic candidate, with demonstrated antiferroelectricity and antiferromagnetism[169]. It has already shown bipolar rectification in a tunable single-phase memristor[170].

### *4.5 Extensions of ferroelectricity*

Functional FE materials also display pyroelectric, catalytic and piezoelectric features that allow for yet another window of modern applications. For instance, surface acoustic waves (SAW) can travel along the surface of a piezoelectric material without losing much energy, and convert sound into electricity as well as interact with other types of waves in the material, such as light, electricity, magnetism, or spin[87]. 2D materials are suited for SAW-controlled optoelectronic devices (Section 4.3) because of their large surface area. The electric fields generated from SAW have been shown to easily modulate the electric transport, light interactions and excitonic properties in 2D $WSe_2$ and BP-$MoS_2$ heterostructures among others[171, 172]. High performing surface acoustic wave (SAW) devices that use 2D FE materials like $LiNbO_3$ have also been reported[173, 174]. By utilizing 2D multiferroics, it may be possible to one day achieve reconfigurable SAW devices[87].

Pyroelectricity refers to the phenomenon of charge release as the polarization intensity changes with temperature. As a result, pyroelectric materials are useful for thermal sensing since their polarization strength varies with temperature. Since the temperature modification can be the result of infrared radiations, the rate of change of the polarization alterations is an estimate of radiation intensity. This has notably been shown in multiple 2D graphene/bulk $LiNbO_3$ photodetectors, where infrared laser spot irradiation caused local ferroelectric polarization of the FE, which generated a thermal current[175]. Therefore, ferroelectric materials have potential for photodetective devices.

The FE state can affect the physical and chemical properties of a material's surface. This is notably the case for catalysis. It has been predicted that the stacking of 2D FEs with out-of-plane polarization could be manipulated to achieve a metallic ferroelectric surface state that can be switched to vary adsorption energies and electrocatalytic activity[176]. The band bending at a ferroelectric/semiconductor interface promotes charge separation and transfer, allowing devices such as FeSFETs to control and be altered by surrounding chemical processes[55].

## 5. CHALLENGES AND OUTLOOKS.

### *5.1. Material preparation.*

Despite the intriguing features of the 2D FE materials/systems, various challenges are encountered in material preparation. Most notably, with the exception of a few material systems (e.g., $In_2Se_3$[37-39, 42], SnS[24, 29, 59] and SnSe[25]), research on 2D ferroelectricity has predominantly utilized small flakes (< a few hundred μm) mechanically exfoliated from a bulk crystal[27, 30, 32-35, 39-41, 44-46, 48, 50, 52, 61]. Given the imperative for wafer-scale electronics applications, there is a pressing need for large-area growth of high quality 2D FE material using bottom-up processes.

Molecular beam epitaxy (MBE) and chemical vapor deposition (CVD) can be used to prepare 2D $In_2Se_3$[37-39, 42] and SnS.[24, 29, 59] and SnSe[25] layers. However, the reported 2D FE flakes are small (< several hundred μm) and polycrystalline. They also lack thickness controllability, which hinders the control of OOP spontaneous polarizations for sliding ferroelectric materials. In addition, defects at the grain boundaries of polycrystalline thin films possess may act as scattering center and undesirable ferroelectric domain walls. MBE is typically employed for research purposes rather than mass production, primarily because of its small throughput when compared to CVD. Therefore, epitaxial growth of single-crystalline 2D ferroelectric material on a crystalline substrate using CVD may be a feasible solution to obtain high quality products.

Furthermore, group III-VI transition metal chalcogenides exhibit different polymorphs

(e.g., α, β, λ phases) or stoichiometries (in the form of MX or $M_2X_3$, where M is a metal and X is a chalcogen atom)[177], requiring precise control for desired FE properties during growth with controlled defects. For example, reports on the general plastic deformation of these material systems have emerged. The ultra-high plasticity is linked to phase transitions, interlayer gliding, and micro-cracks[178-180]. This underscores the significance of designing and producing 2D group III-VI transition metal chalcogenides with controlled defects and layer numbers to achieve desired FE properties during large-area growth. Increasing the conformality is strongly recommended by reducing the growth rate and suppressing the vertical growth mode completely[181].

It is also worth noting that mechanically exfoliated 2D flakes may exhibit different FE properties depending on the material processing and the quality of the mother single crystal. For instance, $CuCrP_2S_6$ prepared by the flux growth exhibits room-temperature ferroelectric ordering[45], while single crystals prepared by the chemical vapor transport displays antiferroelectric properties[46]. These different synthetic approaches for $CuCrP_2S_6$ single crystals could lead to alteration in phase and crystallinity, which likely contributes to the observed variations in the FE phase of $CuCrP_2S_6$. Furthermore, the structural, electrical, and phase instability of $In_2Se_3$ and $WTe_2$ are highly dependent on their oxidation and degradation under air atmosphere and light[182-185]. Besides, Ga vacancy defects, asymmetrically located within the cubic defect semiconductor α-$Ga_2Se_3$, can induce switchable polarizations[186], which further indicates the importance of the defect control.

### *5.2. Reliable characterizations.*

To evaluate ferroelectricity in 2D layers, it is crucial to employ diverse and reliable methods for conformation of polarization properties. When conducting polarization-electric field (P-E) loop measurements, the Sawyer-Tower method is sensitive to interference from various factors, including parasitic, series dielectric, or conducting components (especially leakage currents) in 2D nanodevices[187]. These factors may hinder the isolation of a pure ferroelectric signal especially in 2D semiconductors where metal contacts are also injecting and moving free carriers in the bands as drift current upon application of voltage. As an alternative, piezoresponse force microscopy (PFM) measurements can be used to characterize hysteresis loops. However, discrepancies observed in PFM may not necessarily arise from inherent piezoelectric properties; instead, they could be attributed to electrostatic forces generated by charge deposition on the 2D surface, facilitated by the conducting tip subjected to bias[188]. To address this issue, researchers can conduct measurements using either the double-wave method or the "Positive-Up Negative-Down (PUND)" method to obtain a clear and uncontaminated ferroelectric signal. This is especially important considering the limited effort of such measurements in the evaluation of 2D ferroelectric materials. Nevertheless, the PUND approach may also face complexity due to factors such as leakage currents, varying contact geometries, and small capacitance values[189]. Therefore, it is crucial for researchers to understand the limitations of measurement systems and demonstrate the cross-validation of the existence of switchable polarizations for 2D ferroelectric systems via multiple probes and techniques.

### *5.3. Device fabrication challenges.*

Although FE materials and devices have numerous potential applications (**Section 4.5**), to date their technological impact in high-volume manufacturing products have been limited. This delayed technological impact is in part from the failure points of FE materials (**Section 5.1**). Industry requires device endurances (> $10^9$ cycles) , operation speeds (< 10 ns/transition) and power consumptions (< 10 pJ/transition) beyond what has been reported so far[161, 190]. For instance, a gradual degradation of remanent polarization during cycling is inevitable, as the stress increases trap density near the electrodes, resulting in the domain pinning of dipoles[191]. In high-power applications, energy dissipation in the FE devices ruins performance. Conventional semiconductor channels are not guaranteed to compensate for large remnant polarization values in FeFETs[192]. Aggressive scaling in electronics also requires ferroelectric-based devices to perform at 2D thicknesses, where domain wall pinning is much stronger and meticulous interface engineering is required[193]. In neuromorphic computing, large-scale crossbar arrays of FTJs require the addition of selectors to differentiate the leakage current of the crossbar array, due to the quasi-linear I–V relationships of their ON/OFF states[152]. Non-ideal defects, unwanted chemical reactions, and large serial capacitances at the junctions are a significant burden (**Section 3**). For instance, the unwanted formation of water molecules at the material interface in devices like FTJs and FeFETs can deviate the device's reproducibility and predictability[194, 195]. Moreover, device contacts also pose a challenge as the metal deposition can damage the 2D materials and cause metal-induced gap states that lead to undesired large Schottky barrier heights and contact resistances[193]. This drastically reduces the list of eligible ferroelectric materials for industrial use.

FEs are not always compatible with semiconductor industry constraints, such as fabrication methods, exposure, and growth at temperatures >300°C. Unfortunately, the transition temperature for FE crystals can exceed the operating temperatures of CMOS and common procedures such as hydrogen annealing have a negative impact on FE performance[196]. Financial constraints and availability are also key factors in commercial applications. Ferroelectric rapid access memories (FeRAMs) for instance have lower storage densities than today's DRAMs for a higher price. Extensive use of multiferroics also seems unrealistic, as they are both scarce in nature and costly to synthesize.

## 6. CONCLUSION

2D ferroelectrics and conventional ferroelectrics coupled with 2D structures have rapidly gained prominence in numerous conventional and emerging fields related to electronic materials and devices. In recent years, 2D ferroelectric and semiconductor materials have seen notable improvements in quality, controllability, stability, and growth. The discovery of phenomena such as sliding ferroelectricity and ferroelectric metallicity further underscores their potential for discovery of novel phenomena. However, challenges remain, particularly in terms of reliability and uncertain partial polarizations, hindering widespread adoption of 2D ferroelectric materials and ferroelectrics devices based on 2D materials in commercial applications. Efforts towards optimizing and fabricating FE devices on a large scale are still necessary. In this regard, the exploration and optimization of 2D ferroelectrics for device scaling and next-generation architectures, coupled with systematic reliability considerations, represents a pivotal juncture in the evolution of computing technologies. A pressing problem of integrating 2D semiconductors with conventional ferroelectric materials over wafer scales also remains to be addressed. A clear understanding of the strengths and potential of 2D semiconductor systems, their intrinsic ferroelectric properties as well as their interfaces with 3D ferroelectrics will be important in future device and technology level implementations.

## AUTHOR INFORMATION

**Corresponding Author**

Deep Jariwala

**Authors**

Chloe Leblanc

Seunguk Song

**Author Contributions**

C.L. and S.S. contributed equally to this work.

**Notes**

The authors declare no competing financial interest.

## ACKNOWLEDGMENTS

D.J., S.S. and C.L. acknowledge support from Air Force Office of Scientific Research (AFOSR) GHz-THz program grant number FA9550-23-1-0391 as well as Office of Naval Research (ONR) Nanoscale Computing and Devices program (N00014-24-1-2131). S.S. also acknowledges partial support for this work by Basic Science Research Program through the National Research Foundation of Korea (NRF) funded by the Ministry of Education (Grant No. 2021R1A6A3A14038492).

## REFERENCES

(1) Merolla, P. A.; Arthur, J. V.; Alvarez-Icaza, R.; Cassidy, A. S.; Sawada, J.; Akopyan, F.; Jackson, B. L.; Imam, N.; Guo, C.; Nakamura, Y.; et al. A million spiking-neuron integrated circuit with a scalable communication network and interface. *Science* **2014**, *345* (6197), 668-673. DOI: doi:10.1126/science.1254642.

(2) Indiveri, G.; Liu, S. C. Memory and Information Processing in Neuromorphic Systems. *Proceedings of the Ieee* **2015**, *103* (8), 1379-1397. DOI: 10.1109/jproc.2015.2444094.

(3) Berdan, R.; Marukame, T.; Ota, K.; Yamaguchi, M.; Saitoh, M.; Fujii, S.; Deguchi, J.; Nishi, Y. Low-power linear computation using nonlinear ferroelectric tunnel junction memristors. *Nature Electronics* **2020**, *3* (5), 259-266. DOI: 10.1038/s41928-020-0405-0.

(4) Molas, G.; Nowak, E. Advances in Emerging Memory Technologies: From Data Storage to Artificial Intelligence. *Applied Sciences* **2021**, *11* (23), 11254.

(5) Kang, K.; Xie, S.; Huang, L.; Han, Y.; Huang, P. Y.; Mak, K. F.; Kim, C.-J.; Muller, D.; Park, J. High-mobility three-atom-thick semiconducting films with wafer-scale homogeneity. *Nature* **2015**, *520* (7549), 656-660. DOI: 10.1038/nature14417.

(6) Zhang, K.; She, Y.; Cai, X.; Zhao, M.; Liu, Z.; Ding, C.; Zhang, L.; Zhou, W.; Ma, J.; Liu, H.; et al. Epitaxial substitution of metal iodides for low-temperature growth of two-dimensional metal chalcogenides. *Nature Nanotechnology* **2023**. DOI: 10.1038/s41565-023-01326-1.

(7) Zhu, J.; Park, J.-H.; Vitale, S. A.; Ge, W.; Jung, G. S.; Wang, J.; Mohamed, M.; Zhang, T.; Ashok, M.; Xue, M.; et al. Low-thermal-budget synthesis of monolayer molybdenum disulfide for silicon back-end-of-line integration on a 200 mm platform. *Nature Nanotechnology* **2023**, *18* (5), 456-463. DOI: 10.1038/s41565-023-01375-6.

(8) Samadi, M.; Sarikhani, N.; Zirak, M.; Zhang, H.; Zhang, H.-L.; Moshfegh, A. Z. Group 6 transition

metal dichalcogenide nanomaterials: synthesis, applications and future perspectives. *Nanoscale Horizons* **2018**, *3*. DOI: 10.1039/C7NH00137A.
(9) Wu, Z.; Lyu, Y.; Zhang, Y.; Ding, R.; Zheng, B.; Yang, Z.; Lau, S. P.; Chen, X. H.; Hao, J. Large-scale growth of few-layer two-dimensional black phosphorus. *Nature Materials* **2021**, *20* (9), 1203-1209. DOI: 10.1038/s41563-021-01001-7.
(10) Wang, Y.; Qiu, G.; Wang, R.; Huang, S.; Wang, Q.; Liu, Y.; Du, Y.; Goddard, W. A.; Kim, M. J.; Xu, X.; et al. Field-effect transistors made from solution-grown two-dimensional tellurene. *Nature Electronics* **2018**, *1* (4), 228-236. DOI: 10.1038/s41928-018-0058-4.
(11) Zhao, C.; Tan, C.; Lien, D.-H.; Song, X.; Amani, M.; Hettick, M.; Nyein, H. Y. Y.; Yuan, Z.; Li, L.; Scott, M. C.; et al. Evaporated tellurium thin films for p-type field-effect transistors and circuits. *Nature Nanotechnology* **2020**, *15* (1), 53-58. DOI: 10.1038/s41565-019-0585-9.
(12) Liu, J.; Su, L.; Zhang, X.; Shtansky, D. V.; Fang, X. Ferroelectric-Optoelectronic Hybrid System for Photodetection. *Small Methods* **2023**. DOI: 10.1002/smtd.202300319.
(13) Yasuda, K.; Wang, X.; Watanabe, K.; Taniguchi, T.; Jarillo-Herrero, P. Stacking-engineered ferroelectricity in bilayer boron nitride. *Science* **2021**, *372* (6549), research-article. DOI: 10.1126/science.abd3230.
(14) Iqbal, M. A.; Xie, H.; Qi, L.; Jiang, W.-C.; Zeng, Y.-J. Recent Advances in Ferroelectric-Enhanced Low-Dimensional Optoelectronic Devices. *Small* **2023**, *19* (16). DOI: 10.1002/smll.202205347.
(15) Kim, K.-H.; Oh, S.; Fiagbenu, M. M. A.; Zheng, J.; Musavigharavi, P.; Kumar, P.; Trainor, N.; Aljarb, A.; Wan, Y.; Kim, H. M.; et al. Scalable CMOS back-end-of-line-compatible AlScN/two-dimensional channel ferroelectric field-effect transistors. *Nature Nanotechnology* **2023**, *18* (9), 1044-1050. DOI: 10.1038/s41565-023-01399-y.
(16) Kim, K.-H.; Song, S.; Kim, B.; Musavigharavi, P.; Trainor, N.; Katti, K.; Chen, C.; Kumari, S.; Zheng, J.; Redwing, J. M.; et al. Tuning Polarity in WSe2/AlScN FeFETs via Contact Engineering. *ACS Nano* **2024**, *18* (5), 4180-4188. DOI: 10.1021/acsnano.3c09279.
(17) Kim, K.-H.; Karpov, I.; Olsson, R. H.; Jariwala, D. Wurtzite and fluorite ferroelectric materials for electronic memory. *Nature Nanotechnology* **2023**, *18* (5), 422-441. DOI: 10.1038/s41565-023-01361-y.
(18) Cheema, S. S.; Kwon, D.; Shanker, N.; dos Reis, R.; Hsu, S.-L.; Xiao, J.; Zhang, H.; Wagner, R.; Datar, A.; McCarter, M. R.; et al. Enhanced ferroelectricity in ultrathin films grown directly on silicon. *Nature* **2020**, *580* (7804), 478-482. DOI: 10.1038/s41586-020-2208-x.
(19) Gong, N.; Ma, T. P. Why Is FE–HfO2 More Suitable Than PZT or SBT for Scaled Nonvolatile 1-T Memory Cell? A Retention Perspective. *IEEE Electron Device Letters* **2016**, *37* (9), 1123-1126. DOI: 10.1109/LED.2016.2593627.
(20) Gao, P.; Zhang, Z.; Li, M.; Ishikawa, R.; Feng, B.; Liu, H.-J.; Huang, Y.-L.; Shibata, N.; Ma, X.; Chen, S.; et al. Possible absence of critical thickness and size effect in ultrathin perovskite ferroelectric films. *Nature Communications* **2017**, *8* (1), 15549. DOI: 10.1038/ncomms15549.
(21) Oh, S.; Kim, H.; Kashir, A.; Hwang, H. Effect of dead layers on the ferroelectric property of

ultrathin HfZrOx film. *Applied Physics Letters* **2020**, *117* (25). DOI: 10.1063/5.0030856 (acccessed 1/31/2024).
(22) Sarott, M. F.; Bucheli, U.; Lochmann, A.; Fiebig, M.; Trassin, M. Controlling the Polarization in Ferroelectric PZT Films via the Epitaxial Growth Conditions. *Advanced Functional Materials* **2023**, *33* (28), 2214849. DOI: https://doi.org/10.1002/adfm.202214849.
(23) Paull, O.; Xu, C.; Cheng, X.; Zhang, Y.; Xu, B.; Kelley, K. P.; de Marco, A.; Vasudevan, R. K.; Bellaiche, L.; Nagarajan, V.; et al. Anisotropic epitaxial stabilization of a low-symmetry ferroelectric with enhanced electromechanical response. *Nature Materials* **2022**, *21* (1), 74-80. DOI: 10.1038/s41563-021-01098-w.
(24) Higashitarumizu, N.; Kawamoto, H.; Lee, C.-J.; Lin, B.-H.; Chu, F.-H.; Yonemori, I.; Nishimura, T.; Wakabayashi, K.; Chang, W.-H.; Nagashio, K. Purely in-plane ferroelectricity in monolayer SnS at room temperature. *Nature Communications* **2020**, *11* (1), 2428. DOI: 10.1038/s41467-020-16291-9.
(25) Chang, K.; Küster, F.; Miller, B. J.; Ji, J.-R.; Zhang, J.-L.; Sessi, P.; Barraza-Lopez, S.; Parkin, S. S. P. Microscopic Manipulation of Ferroelectric Domains in SnSe Monolayers at Room Temperature. *Nano Letters* **2020**, *20* (9), 6590-6597. DOI: 10.1021/acs.nanolett.0c02357.
(26) Chang, K.; Liu, J.; Lin, H.; Wang, N.; Zhao, K.; Zhang, A.; Jin, F.; Zhong, Y.; Hu, X.; Duan, W.; et al. Discovery of robust in-plane ferroelectricity in atomic-thick SnTe. *Science* **2016**, *353* (6296), 274-278. DOI: doi:10.1126/science.aad8609.
(27) Xue, F.; Hu, W.; Lee, K.-C.; Lu, L.-S.; Zhang, J.; Tang, H.-L.; Han, A.; Hsu, W.-T.; Tu, S.; Chang, W.-H.; et al. Room-Temperature Ferroelectricity in Hexagonally Layered α-In2Se3 Nanoflakes down to the Monolayer Limit. *Advanced Functional Materials* **2018**, *28* (50), 1803738. DOI: https://doi.org/10.1002/adfm.201803738.
(28) Liu, F.; You, L.; Seyler, K. L.; Li, X.; Yu, P.; Lin, J.; Wang, X.; Zhou, J.; Wang, H.; He, H.; et al. Room-temperature ferroelectricity in CuInP2S6 ultrathin flakes. *Nature Communications* **2016**, *7* (1), 12357. DOI: 10.1038/ncomms12357.
(29) Kwon, K. C.; Zhang, Y.; Wang, L.; Yu, W.; Wang, X.; Park, I.-H.; Choi, H. S.; Ma, T.; Zhu, Z.; Tian, B.; et al. In-Plane Ferroelectric Tin Monosulfide and Its Application in a Ferroelectric Analog Synaptic Device. *ACS Nano* **2020**, *14* (6), 7628-7638. DOI: 10.1021/acsnano.0c03869.
(30) Sharma, P.; Xiang, F.-X.; Shao, D.-F.; Zhang, D.; Tsymbal, E. Y.; Hamilton, A. R.; Seidel, J. A room-temperature ferroelectric semimetal. *Science Advances* **2019**, *5* (7), eaax5080. DOI: doi:10.1126/sciadv.aax5080.
(31) Fei, Z.; Zhao, W.; Palomaki, T. A.; Sun, B.; Miller, M. K.; Zhao, Z.; Yan, J.; Xu, X.; Cobden, D. H. Ferroelectric switching of a two-dimensional metal. *Nature* **2018**, *560* (7718), 336-339. DOI: 10.1038/s41586-018-0336-3.
(32) Niu, R.; Li, Z.; Han, X.; Qu, Z.; Ding, D.; Wang, Z.; Liu, Q.; Liu, T.; Han, C.; Watanabe, K.; et al. Giant ferroelectric polarization in a bilayer graphene heterostructure. *Nature Communications* **2022**, *13* (1), 6241. DOI: 10.1038/s41467-022-34104-z.
(33) Zheng, Z.; Ma, Q.; Bi, Z.; de la Barrera, S.; Liu, M.-H.; Mao, N.; Zhang, Y.; Kiper, N.; Watanabe, K.;

Taniguchi, T.; et al. Unconventional ferroelectricity in moiré heterostructures. *Nature* **2020**, *588* (7836), 71-76. DOI: 10.1038/s41586-020-2970-9.
(34) Yan, X.; Zheng, Z.; Sangwan, V. K.; Qian, J. H.; Wang, X.; Liu, S. E.; Watanabe, K.; Taniguchi, T.; Xu, S.-Y.; Jarillo-Herrero, P.; et al. Moiré synaptic transistor with room-temperature neuromorphic functionality. *Nature* **2023**, *624* (7992), 551-556. DOI: 10.1038/s41586-023-06791-1.
(35) Yan, Y.; Deng, Q.; Li, S.; Guo, T.; Li, X.; Jiang, Y.; Song, X.; Huang, W.; Yang, J.; Xia, C. In-plane ferroelectricity in few-layered GeS and its van der Waals ferroelectric diodes. *Nanoscale* **2021**, *13* (38), 16122-16130, 10.1039/D1NR03807A. DOI: 10.1039/D1NR03807A.
(36) Gou, J.; Bai, H.; Zhang, X.; Huang, Y. L.; Duan, S.; Ariando, A.; Yang, S. A.; Chen, L.; Lu, Y.; Wee, A. T. S. Two-dimensional ferroelectricity in a single-element bismuth monolayer. *Nature* **2023**, *617* (7959), 67-72. DOI: 10.1038/s41586-023-05848-5.
(37) Io, W. F.; Yuan, S.; Pang, S. Y.; Wong, L. W.; Zhao, J.; Hao, J. Temperature- and thickness-dependence of robust out-of-plane ferroelectricity in CVD grown ultrathin van der Waals α-In2Se3 layers. *Nano Research* **2020**, *13* (7), 1897-1902. DOI: 10.1007/s12274-020-2640-0.
(38) Xiao, J.; Zhu, H.; Wang, Y.; Feng, W.; Hu, Y.; Dasgupta, A.; Han, Y.; Wang, Y.; Muller, D. A.; Martin, L. W.; et al. Intrinsic Two-Dimensional Ferroelectricity with Dipole Locking. *Physical Review Letters* **2018**, *120* (22), 227601. DOI: 10.1103/PhysRevLett.120.227601.
(39) Zhou, Y.; Wu, D.; Zhu, Y.; Cho, Y.; He, Q.; Yang, X.; Herrera, K.; Chu, Z.; Han, Y.; Downer, M. C.; et al. Out-of-Plane Piezoelectricity and Ferroelectricity in Layered α-In2Se3 Nanoflakes. *Nano Letters* **2017**, *17* (9), 5508-5513. DOI: 10.1021/acs.nanolett.7b02198.
(40) Parker, J.; Gu, Y. Highly Efficient Polarization-Controlled Electrical Conductance Modulation in a van der Waals Ferroelectric/Semiconductor Heterostructure. *Advanced Electronic Materials* **2022**, *8* (9), 2200413. DOI: https://doi.org/10.1002/aelm.202200413.
(41) Zheng, C.; Yu, L.; Zhu, L.; Collins, J. L.; Kim, D.; Lou, Y.; Xu, C.; Li, M.; Wei, Z.; Zhang, Y.; et al. Room temperature in-plane ferroelectricity in van der Waals In<sub>2</sub>Se<sub>3</sub>. *Science Advances* **2018**, *4* (7), eaar7720. DOI: doi:10.1126/sciadv.aar7720.
(42) Xu, C.; Mao, J.; Guo, X.; Yan, S.; Chen, Y.; Lo, T. W.; Chen, C.; Lei, D.; Luo, X.; Hao, J.; et al. Two-dimensional ferroelasticity in van der Waals β'-In2Se3. *Nature Communications* **2021**, *12* (1), 3665. DOI: 10.1038/s41467-021-23882-7.
(43) Zhang, Z.; Nie, J.; Zhang, Z.; Yuan, Y.; Fu, Y.-S.; Zhang, W. Atomic Visualization and Switching of Ferroelectric Order in β-In2Se3 Films at the Single Layer Limit. *Advanced Materials* **2022**, *34* (3), 2106951. DOI: https://doi.org/10.1002/adma.202106951.
(44) Wan, Y.; Hu, T.; Mao, X.; Fu, J.; Yuan, K.; Song, Y.; Gan, X.; Xu, X.; Xue, M.; Cheng, X.; et al. Room-Temperature Ferroelectricity in $1{₩mathrm{T}}^{₩ensuremath{'}}$-${₩mathrm{ReS}}_{2}$ Multilayers. *Physical Review Letters* **2022**, *128* (6), 067601. DOI: 10.1103/PhysRevLett.128.067601.
(45) Io, W. F.; Pang, S. Y.; Wong, L. W.; Zhao, Y.; Ding, R.; Mao, J.; Zhao, Y.; Guo, F.; Yuan, S.; Zhao, J.; et al. Direct observation of intrinsic room-temperature ferroelectricity in 2D layered CuCrP2S6.

*Nature Communications* **2023**, *14* (1), 7304. DOI: 10.1038/s41467-023-43097-2.
(46) Ma, Y.; Yan, Y.; Luo, L.; Pazos, S.; Zhang, C.; Lv, X.; Chen, M.; Liu, C.; Wang, Y.; Chen, A.; et al. High-performance van der Waals antiferroelectric CuCrP2S6-based memristors. *Nature Communications* **2023**, *14* (1), 7891. DOI: 10.1038/s41467-023-43628-x.
(47) Ma, R.-R.; Xu, D.-D.; Zhong, Q.-L.; Zhong, C.-R.; Huang, R.; Xiang, P.-H.; Zhong, N.; Duan, C.-G. Nanoscale Mapping of Cu-Ion Transport in van der Waals Layered CuCrP2S6. *Advanced Materials Interfaces* **2022**, *9* (4), 2101769. DOI: https://doi.org/10.1002/admi.202101769.
(48) Wang, X.; Zhu, C.; Deng, Y.; Duan, R.; Chen, J.; Zeng, Q.; Zhou, J.; Fu, Q.; You, L.; Liu, S.; et al. Van der Waals engineering of ferroelectric heterostructures for long-retention memory. *Nature Communications* **2021**, *12* (1), 1109. DOI: 10.1038/s41467-021-21320-2.
(49) Studenyak, I. P.; Mitrovcij, V. V.; Kovacs, G. S.; Gurzan, M. I.; Mykajlo, O. A.; Vysochanskii, Y. M.; Cajipe, V. B. Disordering effect on optical absorption processes in CuInP2S6 layered ferrielectrics. *physica status solidi (b)* **2003**, *236* (3), 678-686. DOI: https://doi.org/10.1002/pssb.200301513.
(50) Deb, S.; Cao, W.; Raab, N.; Watanabe, K.; Taniguchi, T.; Goldstein, M.; Kronik, L.; Urbakh, M.; Hod, O.; Ben Shalom, M. Cumulative polarization in conductive interfacial ferroelectrics. *Nature* **2022**, *612* (7940), 465-469. DOI: 10.1038/s41586-022-05341-5.
(51) Yang, T. H.; Liang, B.-W.; Hu, H.-C.; Chen, F.-X.; Ho, S.-Z.; Chang, W.-H.; Yang, L.; Lo, H.-C.; Kuo, T.-H.; Chen, J.-H.; et al. Ferroelectric transistors based on shear-transformation-mediated rhombohedral-stacked molybdenum disulfide. *Nature Electronics* **2023**. DOI: 10.1038/s41928-023-01073-0.
(52) Wang, X.; Yasuda, K.; Zhang, Y.; Liu, S.; Watanabe, K.; Taniguchi, T.; Hone, J.; Fu, L.; Jarillo-Herrero, P. Interfacial ferroelectricity in rhombohedral-stacked bilayer transition metal dichalcogenides. *Nature Nanotechnology* **2022**, *17* (4), 367-371. DOI: 10.1038/s41565-021-01059-z.
(53) McCreary, K. M.; Phillips, M.; Chuang, H.-J.; Wickramaratne, D.; Rosenberger, M.; Hellberg, C. S.; Jonker, B. T. Stacking-dependent optical properties in bilayer WSe2. *Nanoscale* **2022**, *14* (1), 147-156, 10.1039/D1NR06119D. DOI: 10.1039/D1NR06119D.
(54) Song, S.; Sim, Y.; Kim, S.-Y.; Kim, J. H.; Oh, I.; Na, W.; Lee, D. H.; Wang, J.; Yan, S.; Liu, Y.; et al. Wafer-scale production of patterned transition metal ditelluride layers for two-dimensional metal–semiconductor contacts at the Schottky–Mott limit. *Nature Electronics* **2020**, *3* (4), 207-215. DOI: 10.1038/s41928-020-0396-x.
(55) Song, S.; Yoon, A.; Jang, S.; Lynch, J.; Yang, J.; Han, J.; Choe, M.; Jin, Y. H.; Chen, C. Y.; Cheon, Y.; et al. Fabrication of p-type 2D single-crystalline transistor arrays with Fermi-level-tuned van der Waals semimetal electrodes. *Nature Communications* **2023**, *14* (1), 4747. DOI: 10.1038/s41467-023-40448-x.
(56) Lyu, F.; Sun, Y.; Yang, Q.; Tang, B.; Li, M.; Li, Z.; Sun, M.; Gao, P.; Ye, L.-H.; Chen, Q. Thickness-dependent band gap of α-In2Se3: from electron energy loss spectroscopy to density functional theory calculations. *Nanotechnology* **2020**, *31* (31), 315711. DOI: 10.1088/1361-6528/ab8998.
(57) Wan, S.; Li, Y.; Li, W.; Mao, X.; Wang, C.; Chen, C.; Dong, J.; Nie, A.; Xiang, J.; Liu, Z.; et al.

Nonvolatile Ferroelectric Memory Effect in Ultrathin α-In2Se3. *Advanced Functional Materials* **2019**, *29* (20), 1808606. DOI: https://doi.org/10.1002/adfm.201808606.
(58) Fei, R.; Kang, W.; Yang, L. Ferroelectricity and Phase Transitions in Monolayer Group-IV Monochalcogenides. *Physical Review Letters* **2016**, *117* (9), 097601. DOI: 10.1103/PhysRevLett.117.097601.
(59) Bao, Y.; Song, P.; Liu, Y.; Chen, Z.; Zhu, M.; Abdelwahab, I.; Su, J.; Fu, W.; Chi, X.; Yu, W.; et al. Gate-Tunable In-Plane Ferroelectricity in Few-Layer SnS. *Nano Letters* **2019**, *19* (8), 5109-5117. DOI: 10.1021/acs.nanolett.9b01419.
(60) Zhu, L.; Lu, Y.; Wang, L. Tuning ferroelectricity by charge doping in two-dimensional SnSe. *Journal of Applied Physics* **2020**, *127* (1). DOI: 10.1063/1.5123296 (acccessed 1/22/2024).
(61) Guan, Z.; Zhao, Y.; Wang, X.; Zhong, N.; Deng, X.; Zheng, Y.; Wang, J.; Xu, D.; Ma, R.; Yue, F.; et al. Electric-Field-Induced Room-Temperature Antiferroelectric–Ferroelectric Phase Transition in van der Waals Layered GeSe. *ACS Nano* **2022**, *16* (1), 1308-1317. DOI: 10.1021/acsnano.1c09183.
(62) Hu, Y.; Zhang, S.; Sun, S.; Xie, M.; Cai, B.; Zeng, H. GeSe monolayer semiconductor with tunable direct band gap and small carrier effective mass. *Applied Physics Letters* **2015**, *107* (12). DOI: 10.1063/1.4931459 (acccessed 1/22/2024).
(63) Xue, F.; He, X.; Retamal, J. R. D.; Han, A.; Zhang, J.; Liu, Z.; Huang, J.-K.; Hu, W.; Tung, V.; He, J.-H.; et al. Gate-Tunable and Multidirection-Switchable Memristive Phenomena in a Van Der Waals Ferroelectric. *Advanced Materials* **2019**, *31* (29), 1901300. DOI: https://doi.org/10.1002/adma.201901300.
(64) Jiang, X.; Feng, Y.; Chen, K.-Q.; Tang, L.-M. The coexistence of ferroelectricity and topological phase transition in monolayer α-In2Se3 under strain engineering. *Journal of Physics: Condensed Matter* **2020**, *32* (10), 105501. DOI: 10.1088/1361-648X/ab58f1.
(65) Jin, X.; Zhang, Y.-Y.; Du, S. Recent progress in the theoretical design of two-dimensional ferroelectric materials. *Fundamental Research* **2023**, *3* (3), 322-331. DOI: https://doi.org/10.1016/j.fmre.2023.02.009.
(66) Ding, W.; Zhu, J.; Wang, Z.; Gao, Y.; Xiao, D.; Gu, Y.; Zhang, Z.; Zhu, W. Prediction of intrinsic two-dimensional ferroelectrics in In2Se3 and other III2-VI3 van der Waals materials. *Nature Communications* **2017**, *8* (1), 14956. DOI: 10.1038/ncomms14956.
(67) Hu, W. J.; Juo, D.-M.; You, L.; Wang, J.; Chen, Y.-C.; Chu, Y.-H.; Wu, T. Universal Ferroelectric Switching Dynamics of Vinylidene Fluoride-trifluoroethylene Copolymer Films. *Scientific Reports* **2014**, *4* (1), 4772. DOI: 10.1038/srep04772.
(68) Xu, C.; Chen, Y.; Cai, X.; Meingast, A.; Guo, X.; Wang, F.; Lin, Z.; Lo, T. W.; Maunders, C.; Lazar, S.; et al. Two-Dimensional Antiferroelectricity in Nanostripe-Ordered ${₩mathrm{In}}_{2}{₩mathrm{Se}}_{3}$. *Physical Review Letters* **2020**, *125* (4), 047601. DOI: 10.1103/PhysRevLett.125.047601.
(69) Li, W.; Zhang, X.; Yang, J.; Zhou, S.; Song, C.; Cheng, P.; Zhang, Y.-Q.; Feng, B.; Wang, Z.; Lu, Y.; et al. Emergence of ferroelectricity in a nonferroelectric monolayer. *Nature Communications* **2023**, *14*

(1), 2757. DOI: 10.1038/s41467-023-38445-1.
(70) Du, R.; Wang, Y.; Cheng, M.; Wang, P.; Li, H.; Feng, W.; Song, L.; Shi, J.; He, J. Two-dimensional multiferroic material of metallic p-doped SnSe. *Nature Communications* **2022**, *13* (1), 6130. DOI: 10.1038/s41467-022-33917-2.
(71) Singh, P.; Baek, S.; Yoo, H. H.; Niu, J.; Park, J.-H.; Lee, S. Two-Dimensional CIPS-InSe van der Waal Heterostructure Ferroelectric Field Effect Transistor for Nonvolatile Memory Applications. *ACS Nano* **2022**, *16* (4), 5418-5426. DOI: 10.1021/acsnano.1c09136.
(72) Ram, A.; Maity, K.; Marchand, C.; Mahmoudi, A.; Kshirsagar, A. R.; Soliman, M.; Taniguchi, T.; Watanabe, K.; Doudin, B.; Ouerghi, A.; et al. Reconfigurable Multifunctional van der Waals Ferroelectric Devices and Logic Circuits. *ACS Nano* **2023**, *17* (21), 21865-21877. DOI: 10.1021/acsnano.3c07952.
(73) Vizner Stern, M.; Waschitz, Y.; Cao, W.; Nevo, I.; Watanabe, K.; Taniguchi, T.; Sela, E.; Urbakh, M.; Hod, O.; Ben Shalom, M. Interfacial ferroelectricity by van der Waals sliding. *Science* **2021**, *372* (6549), 1462-1466. DOI: doi:10.1126/science.abe8177.
(74) Yang, L.; Wu, M. Across-Layer Sliding Ferroelectricity in 2D Heterolayers. *Advanced Functional Materials* **2023**, *33* (29), 2301105. DOI: https://doi.org/10.1002/adfm.202301105.
(75) Sui, F.; Jin, M.; Zhang, Y.; Qi, R.; Wu, Y.-N.; Huang, R.; Yue, F.; Chu, J. Sliding ferroelectricity in van der Waals layered γ-InSe semiconductor. *Nature Communications* **2023**, *14* (1), 36. DOI: 10.1038/s41467-022-35490-0.
(76) Rogée, L.; Wang, L.; Zhang, Y.; Cai, S.; Wang, P.; Chhowalla, M.; Ji, W.; Lau, S. P. Ferroelectricity in untwisted heterobilayers of transition metal dichalcogenides. *Science* **2022**, *376* (6596), 973-978. DOI: doi:10.1126/science.abm5734.
(77) Weston, A.; Castanon, E. G.; Enaldiev, V.; Ferreira, F.; Bhattacharjee, S.; Xu, S.; Corte-León, H.; Wu, Z.; Clark, N.; Summerfield, A.; et al. Interfacial ferroelectricity in marginally twisted 2D semiconductors. *Nature Nanotechnology* **2022**, *17* (4), 390-395. DOI: 10.1038/s41565-022-01072-w.
(78) Jin, T.; Mao, J.; Gao, J.; Han, C.; Loh, K. P.; Wee, A. T. S.; Chen, W. Ferroelectrics-Integrated Two-Dimensional Devices toward Next-Generation Electronics. *ACS Nano* **2022**, *16* (9), 13595–13611, review-article. DOI: 10.1021/acsnano.2c07281.
(79) Yin, L.; Cheng, R.; Liu, C.; He, J. Emerging 2D memory devices for in-memory computing. *Advanced Materials* **2021**, *33* (29). DOI: 10.1002/adma.202007081.
(80) Bailey, B. Von Neumann Is Struggling. *Semiconductor Engineering*, 2021-01-18, 2021. https://semiengineering.com/von-neumann-is-struggling/.
(81) Mikolajick, T.; Schroeder, U.; Slesazeck, S. The Case for Ferroelectrics in Future Memory Devices. In 2021 5th IEEE Electron Devices Technology & Manufacturing Conference (EDTM), 2021.
(82) Wang, D.; Musavigharavi, P.; Zheng, J.; Esteves, G.; Liu, X.; Fiagbenu, M. M. A.; Stach, E. A.; Jariwala, D.; Olsson, R. H. I. Sub-Microsecond Polarization Switching in (Al,Sc)N Ferroelectric Capacitors Grown on Complementary Metal–Oxide–Semiconductor-Compatible Aluminum Electrodes. *Rapid Research Letters* **2021**. DOI: 10.1002/pssr.202000575.

(83) Velev, J. P.; Burton, J. D.; Zhuravlev, M. Y.; Tsymbal, E. Y. Predictive modelling of ferroelectric tunnel junctions. *npj Computational Materials* **2016**, *2* (1), 1-13, ReviewPaper. DOI: 10.1038/npjcompumats.2016.9.
(84) Tsymbal, E. Y.; Velev, J. P. Ferroelectric Tunnel Junctions: Crossing the wall. *Nature Nanotechnology* **2017**, *12* (7), 614-615, BriefCommunication. DOI: 10.1038/nnano.2017.60.
(85) Su, J.; Zheng, X.; Zheng Wen, T. L.; Xie, S.; Rabe, K. M.; Liu, X.; Tsymbal, E. Y. Resonant band engineering of ferroelectric tunnel junctions. *Physical Review B* **2021**, *104* (6). DOI: 10.1103/PhysRevB.104.L060101.
(86) Stengel, M.; Vanderbilt, D.; Spaldin, N. A. Enhancement of ferroelectricity at metal–oxide interfaces. *Nature Materials* **2009**, *8* (5), 392-397, OriginalPaper. DOI: 10.1038/nmat2429.
(87) Luca, G. M. D.; Rubano, A. Two-Dimensional Ferroelectrics: A Review on Applications and Devices. *Solids* **2024**, *5* (1), 45-65, Review. DOI: 10.3390/solids5010004.
(88) Li, Y.; Gong, M.; Zeng, H. Atomically thin α-In2Se3: an emergent two-dimensional room temperature ferroelectric semiconductor. *Journal of Semiconductors* **2019**, *40* (6), Text. DOI: 10.1088/1674-4926/40/6/061002.
(89) Liu, Y.; Wu, Y.; Wang, B.; Chen, H.; Yi, D.; Liu, K.; Nan, C.-W.; Ma, J. Versatile memristor implemented in van der Waals CuInP2S6. *Nano Research* **2023**, *16* (7), 10191-10197, OriginalPaper. DOI: 10.1007/s12274-023-5583-4.
(90) W, L.; Y, G.; Z, L.; S, W.; B, H.; W, H.; L, Y.; K, W.; T, T.; T, A.; et al. A Gate Programmable van der Waals Metal-Ferroelectric-Semiconductor Vertical Heterojunction Memory. *Advanced materials* **2023**, *35* (5). DOI: 10.1002/adma.202208266.
(91) Ma, C.; Luo, Z.; Huang, W.; Zhao, L.; Chen, Q.; Lin, Y.; Liu, X.; Chen, Z.; Liu, C.; Sun, H.; et al. Sub-nanosecond memristor based on ferroelectric tunnel junction. *Nature Communications* **2020**, *11* (1), 1-9, OriginalPaper. DOI: 10.1038/s41467-020-15249-1.
(92) Xi, Z.; Ruan, J.; Li, C.; Zheng, C.; Wen, Z.; Dai, J.; Li, A.; Wu, D. Giant tunnelling electroresistance in metal/ferroelectric/semiconductor tunnel junctions by engineering the Schottky barrier. *Nature Communications* **2017**, *8*, OriginalPaper. DOI: 10.1038/ncomms15217.
(93) Lei, P.; Duan, H.; Qin, L.; Wei, X.; Tao, R.; Wang, Z.; Guo, F.; Song, M.; Jie, W.; Hao, J. **High-Performance Memristor Based on 2D Layered BiOI Nanosheet for Low-Power Artificial Optoelectronic Synapses**. *Advanced Functional Materials* **2022**, *32* (25). DOI: 10.1002/adfm.202201276.
(94) Zhang, M.; Qin, Q.; Chen, X.; Tang, R.; Han, A., Yao; Suhao; Dan, R. Towards an universal artificial synapse using MXene-PZT based ferroelectric memristor. *Ceramics International* **2022**, *48* (11), 16263-16272. DOI: 10.1016/j.ceramint.2022.02.175.
(95) Lu, X. F.; Zhang, Y.; Wang, N.; Luo, S.; Peng, K.; Wang, L.; Chen, H.; Gao, W.; Chen, X. H.; Bao, Y.; et al. Exploring Low Power and Ultrafast Memristor on p-Type van der Waals SnS. *Nano Letters* **2021**, *21* (20), 8800-8807, rapid-communication. DOI: 10.1021/acs.nanolett.1c03169.
(96) Boni, G. A. I., C.M. Zacharaki, C. Tsipas, P. Chaitoglou, S. Evangelou, E.K. Dimoulas, A. Pintilie,

I. Pintilie, L. The Role of Interface Defect States in n- and p-Type Ge Metal–Ferroelectric–Semiconductor Structures with Hf $_{0.5}$ Zr $_{0.5}$ O 2 Ferroelectric. *Phys. Status Silidi A* **2021**, *218*. DOI: 10.1002/pssa.202000500.
(97) Okuno, J.; Kunihiro, T.; Konishi, K.; Materano, M.; Ali, T.; Kuehnel, K.; Seidel, K.; Mikolajick, T.; Schroeder, U.; Tsukamoto, M.; et al. 1T1C FeRAM Memory Array Based on Ferroelectric HZO With Capacitor Under Bitline. *IEEE Journal of the Electron Devices Society* **2021**, *10*, 29-34. DOI: 10.1109/JEDS.2021.3129279.
(98) Jiang, H.; Bhuiyan, M. A.; Liu, Z.; Ma, T. P. A Study of BEOL Processed Hf0.5Zr0.5O2-based Ferroelectric Capacitors and Their Potential for Automotive Applications. In 2020 IEEE International Memory Workshop (IMW), Dresden, Germany; 2020.
(99) Liu, X.; Zheng, J.; logo, D. W. O.; logo, P. M. O.; logo, E. A. S. O.; Roy Olsson, I.; Jariwala, D. Aluminum scandium nitride-based metal–ferroelectric–metal diode memory devices with high on/off ratios. *Applied Physics Letters* **2021**, *118* (20). DOI: 10.1063/5.0051940.
(100) Cheema, S. S.; Shanker, N.; Hsu, C.-H.; Datar, A.; Bae, J.; Kwon, D.; Salahuddin, S. One nanometer HfO2-based ferroelectric tunnel junctions

on silicon. *arXiv* **2020**. DOI: arXiv:2007.06182v1.
(101) Waser, R. *Nanoelectronics and Information Technology*, Wiley-VCH, 2005.
(102) Ajayan, J.; Mohankumar, P.; Nirmal, D.; Leo Joseph, L. M. I.; Bhattacharya, S.; Sreejith, S.; Kollem, S.; Rebelli, S.; Tayal, S.; Mounika, B. Ferroelectric Field-Effect Transistors (FeFETs): Advancements, challenges and exciting prospects for next-generation Non-Volatile Memory (NVM) applications. *Materials Today Communications* **2023**, *35*. DOI: 10.1016/j.mtcomm.2023.105591.
(103) Park, M. H.; Kwon, D.; Schroeder, U.; Mikolajick, T. Binary ferroelectric oxides for future computing paradigms. *MRS Bulletin* **2021**, *46* (11), 1071-1079, ReviewPaper. DOI: 10.1557/s43577-021-00210-4.
(104) Yang, K.; Kim, S. H.; Jeong, H. W.; Lee, D. H.; Park, G. H.; Lee, Y.; Park, M. H. Perspective on Ferroelectric Devices: Lessons from Interfacial Chemistry†. **2023**, review-article. DOI: 10.1021/acs.chemmater.2c03379.
(105) Jiang, X.; Hu, X.; Bian, J.; Zhang, K.; Chen, L.; Zhu, H.; Sun, Q.; Zhang, D. W. Ferroelectric Field-Effect Transistors Based on WSe2/CuInP2S6 Heterostructures for Memory Applications. *ACS Applied Electronic Materials* **2021**, *3* (11), 4711-4717, research-article. DOI: 10.1021/acsaelm.1c00492.
(106) Si, M.; Saha, A. K.; Gao, S.; Qiu, G.; Qin, J.; Duan, Y.; Jian, J.; Niu, C.; Wang, H.; Wu, W.; et al. A ferroelectric semiconductor field-effect transistor. *Nature Electronics* **2019**, *2* (12), 580-586, OriginalPaper. DOI: 10.1038/s41928-019-0338-7.
(107) Liu, M.; Liao, T.; Sun, Z.; Gu, Y.; Kou, L. 2D ferroelectric devices: working principles and research progress. *Physical Chemistry Chemical Physics* **2021**, *23* (38). DOI: 10.1039/D1CP02788C.
(108) Liao, J.; Wen, W.; Wu, J.; Zhou, Y.; Hussain, S.; Hu, H.; Li, J.; Liaqat, A.; Zhu, H.; Jiao, L.; et al. Van der Waals Ferroelectric Semiconductor Field Effect Transistor for In-Memory Computing. *ACS Nano*

**2023**, *17* (6), 6095–6102, research-article. DOI: 10.1021/acsnano.3c01198.
(109) Yang, J. Y.; Yeom, M. J.; Park, Y. o.; Heo, J.; Yoo, G. Ferroelectric α-$In_2Se_3$ Wrapped-Gate β-$Ga_2O_3$ Field-Effect Transistors for Dynamic Threshold Voltage Control. *Advanced Electronic Materials* **2021**, *7* (8). DOI: 10.1002/aelm.202100306.
(110) Alam, M. A.; Si, M.; Ye, P. D. A critical review of recent progress on negative capacitance field-effect transistors. *Applied Physics Letters* **2019**, *114* (9). DOI: 10.1063/1.5092684.
(111) Khan, A. I. Energy-efficient computing with negative capacitance. In *Advanced nanoelectronics : post-silicon*

*materials and devices*, Hussain, M. M. Ed.; Wiley-VCH, 2018; pp 179-196.
(112) McGuire, F. A.; Cheng, Z.; Price, K.; Franklin, A. D. Sub-60 mV/decade switching in 2D negative capacitance field-effect transistors with integrated ferroelectric polymer. *Applied Physics Letters* **2024**, *109*. DOI: 10.1063/1.4961108.
(113) Song, S.; Kim, K.-H.; Chakravarthi, S.; Han, Z.; Kim, G.; Ma, K. Y.; Shin, H. S.; Olsson, R. H., III; Jariwala, D. Negative capacitance field-effect transistors based on ferroelectric AlScN and 2D MoS2. *Applied Physics Letters* **2023**, *123* (18). DOI: 10.1063/5.0169689 (acccessed 2/6/2024).
(114) McGuire, F. A.; Lin, Y.-C.; Price, K.; Rayner, G. B.; Khandelwal, S.; Salahuddin, S.; Franklin, A. D. Sustained Sub-60 mV/decade Switching via the Negative Capacitance Effect in MoS2 Transistors. *Nano Letters* **2017**, *17* (8), 4801-4806, rapid-communication. DOI: 10.1021/acs.nanolett.7b01584.
(115) Si, M.; Su, C.-J.; Jiang, C.; Conrad, N. J.; Zhou, H.; Maize, K. D.; Qiu, G.; Wu, C.-T.; Shakouri, A.; Alam, M. A.; et al. Steep-slope hysteresis-free negative capacitance MoS2 transistors. *Nature Nanotechnology* **2017**, *13* (1), 24-28, OriginalPaper. DOI: doi:10.1038/s41565-017-0010-1.
(116) Khan, A. I.; Chatterjee, K.; Duarte, J. P.; Lu, Z.; Sachid, A.; Khandelwal, S.; Ramesh, R.; Hu, C.; Salahuddin, S. Negative Capacitance in Short-Channel FinFETs Externally Connected to an Epitaxial Ferroelectric Capacitor. *IEEE Electron Device Letters* **2016**, *37* (1), 111-114. DOI: 10.1109/LED.2015.2501319.
(117) Ko, E.; Shin, J.; Shin, C. Steep switching devices for low power applications: negative differential capacitance/resistance field effect transistors. *Nano Convergence* **2018**, *5* (1), ReviewPaper. DOI: 10.1186/s40580-018-0135-4.
(118) Zagni, N.; Alam, M. A. Reliability physics of ferroelectric/negative capacitance transistors for memory/logic applications: An integrative perspective. *Journal of Materials Research* **2021**, *36* (24), 4908-4918. DOI: 10.1557/s43578-021-00420-1.
(119) Ni, K.; Sharma, P.; Zhang, J.; Jerry, M.; Smith, J. A.; Tapily, K.; Clark, R.; Mahapatra, S.; Datta, S. Critical Role of Interlayer in Hf0.5Zr0.5O2 Ferroelectric FET Nonvolatile Memory Performance. *IEEE Transactions on Electron Devices* **2018**, *65* (6), 2461 - 2469. DOI: 10.1109/TED.2018.2829122.
(120) Yoon, S.-J.; Min, D.-H.; Moon, S.-E.; Park, K. S.; Won, J. I.; Yoon, S.-M. Improvement in Long-Term and High-Temperature Retention Stability of Ferroelectric Field-Effect Memory Transistors With Metal–Ferroelectric–Metal–Insulator–Semiconductor Gate-Stacks Using Al-Doped HfO2 Thin Films.

*IEEE Transactions on Electron Devices* **2020**, *67* (2), 499-504. DOI: 10.1109/TED.2019.2961117.
(121) Chatterjee, K.; Kim, S.; Karbasian, G.; Tan, A. J.; Yadav, A. K.; Khan, A. I.; Hu, C.; Salahuddin, S. Self-Aligned, Gate Last, FDSOI, Ferroelectric Gate Memory Device With 5.5-nm Hf0.8 Zr0.2 O2 , High Endurance and Breakdown Recovery. *IEEE Electron Device Letters* **2017**, *38* (10). DOI: 10.1109/LED.2017.2748992.
(122) Mulaosmanovic, H.; Ocker, J.; Müller, S.; Schroeder, U.; Müller, J.; Polakowski, P.; Flachowsky, S.; Bentum, R. v.; Mikolajick, T.; Slesazeck, S. Switching Kinetics in Nanoscale Hafnium Oxide Based Ferroelectric Field-Effect Transistors. *ACS App. Mater. Interfaces* **2017**, *9* (4), 3792–3798, research-article. DOI: 10.1021/acsami.6b13866.
(123) Dünkel, S.; Trentzsch, M.; Richter, R.; Moll, P.; Fuchs, C.; Gehring, O.; Majer, M.; Wittek, S.; Müller, B.; Melde, T.; et al. A FeFET based super-low-power ultra-fast embedded NVM technology for 22nm FDSOI and beyond. In IEEE International Electron Devices Meeting (IEDM), San Francisco; 2017.
(124) Ko, C.; Lee, Y.; Chen, Y.; Suh, J.; Fu, D.; Suslu, A.; Lee, S.; Clarkson, J. D.; Choe, H. S.; Tongay, S.; et al. Ferroelectrically Gated Atomically Thin Transition-Metal Dichalcogenides as Nonvolatile Memory. *Advanced Materials* **2016**, *28* (15). DOI: 10.1002/adma.201504779.
(125) Lipatov, A.; Sharma, P.; Gruverman, A.; Sinitskii, A. Optoelectrical Molybdenum Disulfide (MoS2)—Ferroelectric Memories. *ACS Nano* **2015**, *9* (8), 8089–8098, research-article. DOI: 10.1021/acsnano.5b02078.
(126) Lee, H. S.; Min, S.-W.; Park, M. K.; Lee, Y. T.; Jeon, P. J.; Kim, J. H.; Ryu, S.; Im, S. $MoS_2$ Nanosheets for Top-Gate Nonvolatile Memory Transistor Channel. *Small* **2012**, *8* (20), 3111-3115. DOI: 10.1002/smll.201200752.
(127) Lee, Y. T.; Kwon, H.; Kim, J. S.; Kim, H.-H.; Lee, Y. J.; Lim, J. A.; Song, Y.-W.; Yi, Y.; Choi, W.-K.; Hwang, D. K.; et al. Nonvolatile Ferroelectric Memory Circuit Using Black Phosphorus Nanosheet-Based Field-Effect Transistors with P(VDF-TrFE) Polymer. *ACS Nano* **2015**, *9* (10), 10394–10401, research-article. DOI: 10.1021/acsnano.5b04592.
(128) Wang, X.; Liu, C.; Chen, Y.; Wu, G.; Yan, X.; Huang, H.; Wang, P.; Tian, B.; Hong, Z.; Wang, Y.; et al. Ferroelectric FET for nonvolatile memory application with two-dimensional MoSe2 channels - IOPscience. *2D Materials* **2017**, *4* (2), Text. DOI: 10.1088/2053-1583/aa5c17.
(129) Su, M.; Yang, Z.; Liao, L.; Zou, X.; Ho, J. C.; Wang, J.; Wang, J.; Hu, W.; Xiao, X.; Jiang, C.; et al. Side-Gated $In_2O_3$ Nanowire Ferroelectric FETs for High-Performance Nonvolatile Memory Applications. *Advanced Science* **2016**, *3* (9). DOI: 10.1002/advs.201600078.
(130) Xu, M.; Qi, W.; Li, S.; Wang, W. High-Mobility, Low-Voltage Programmable/Erasable Ferroelectric Polymer Transistor Nonvolatile Memory Based on a P(VDF-TrFE)/PMMA Bilayer Gate Insulator. *IEEE Transactions on Electron Devices* **2021**, *68* (7), 3359-3364. DOI: 10.1109/TED.2021.3077199.
(131) Yang, J. Y.; Park, M.; Yeom, M. J.; Baek, Y.; Yoon, S. C.; Jeong, Y. J.; Oh, S. Y.; Lee, K.; Yoo, G. Reconfigurable Physical Reservoir in GaN/α-In2Se3 HEMTs Enabled by Out-of-Plane Local Polarization of Ferroelectric 2D Layer. *ACS Nano* **2023**, *17* (8), 7695-7704, research-article. DOI: 10.1021/acsnano.3c00187.

(132) Huang, W.; Wang, F.; Yin, L.; Cheng, R.; Wang, Z.; Sendeku, M. G.; Wang, J.; Li, N.; Yao, Y.; He, J. Gate-Coupling-Enabled Robust Hysteresis for Nonvolatile Memory and Programmable Rectifier in Van der Waals Ferroelectric Heterojunctions. *Advanced Materials* **2020**, *32* (14). DOI: 10.1002/adma.201908040.
(133) Si, M.; Liao, P.-Y.; Qiu, G.; Duan, Y.; Ye, P. D. Ferroelectric Field-Effect Transistors Based on MoS2 and CuInP2S6 Two-Dimensional van der Waals Heterostructure. *ACS Publications* **2018**, *12* (7), 6700-6705, research-article. DOI: 10.1021/acsnano.8b01810.
(134) Wan, S.; Li, Y.; Li, W.; Mao, X.; Wang, C.; Chen, C.; Dong, J.; Nie, A.; Xiang, J.; Liu, Z.; et al. Nonvolatile Ferroelectric Memory Effect in Ultrathin α-$In_2Se_3$. *Advanced Functional Materials* **2019**, *29* (20). DOI: 10.1002/adfm.201808606.
(135) Wang, X.; Feng, Z.; Cai, J.; Tong, H.; Miao, X. All-van der Waals stacking ferroelectric field-effect transistor based on In2Se3 for high-density memory. *Science China Information Sciences* **2023**, *66* (8), 1-8, OriginalPaper. DOI: 10.1007/s11432-022-3617-2.
(136) Huo, J.; Zhang, Z.; Zhang, Y.; Zhang, F.; Yan, G.; Tian, G.; Xu, H.; Zhan, G.; Xu, G.; Zhang, Q.; et al. Stacked HZO/α-In2Se2 Ferroelectric Dielectric/Semiconductor FET With Ultrahigh Speed and Large Memory Window. *IEEE Transactions on Electron Devices* **2023**, *70* (6), 3071-3075. DOI: 10.1109/TED.2023.3269403.
(137) Kim, K.; Oh, S.; Fiagbenu, M. M. A.; Zheng, J.; Musavigharavi, P.; Kumar, P.; Trainor, N.; Aljarb, A.; Wan, Y.; Kim, H. M.; et al. Scalable CMOS back-end-of-line-compatible AlScN/two-dimensional channel ferroelectric field-effect transistors. *Nature nanotechnology* **2022**, *18* (9). DOI: 10.1038/s41565-023-01399-y.
(138) Mondal, S.; Wang, D.; Liu, J.; Xiao, Y.; Wang, P.; Mi, Z. ScAlN Based Ferroelectric Field Effect Transistors with ITO Channel. In 2023 Device Research Conference, Santa Barbara, CA; 2023.
(139) Baek, S.; Yoo, H. H.; Ju, J. H.; Sriboriboon, P.; Singh, P.; Niu, J.; Park, J.-H.; Shin, C.; Kim, Y.; Lee, S. Ferroelectric Field-Effect-Transistor Integrated with Ferroelectrics Heterostructure. *Advanced Science* **2022**, *9* (21). DOI: 10.1002/advs.202200566.
(140) Wang, X.; Zhu, C.; Deng, Y.; Duan, R.; Chen, J.; Zeng, Q.; Zhou, J.; Fu, Q.; You, L.; Liu, S.; et al. Van der Waals engineering of ferroelectric heterostructures for long-retention memory. *Nature Communications* **2021**, *12* (1), 1-8, OriginalPaper. DOI: 10.1038/s41467-021-21320-2.
(141) Zha, J.; Shi, S.; Chaturvedi, A.; Huang, H.; Yang, P.; Yao, Y.; Li, S.; Xia, Y.; Zhang, Z.; Wang, W.; et al. Electronic/Optoelectronic Memory Device Enabled by Tellurium-based 2D van der Waals Heterostructure for in-Sensor Reservoir Computing at the Optical Communication Band. *Advanced materials (Deerfield Beach, Fla.)* **2023**, *35* (20). DOI: 10.1002/adma.202211598.
(142) Chen, Y.; Li, D.; Ren, H.; Tang, Y.; Liang, K.; Wang, Y.; Li, F.; Song, C.; Guan, J.; Chen, Z.; et al. Highly Linear and Symmetric Synaptic Memtransistors Based on

Polarization Switching in Two-Dimensional Ferroelectric Semiconductors. *Small* **2022**, *18* (45). DOI: 10.1002/smll.202203611.

(143) Wang, L.; Wang, X.; Zhang, Y.; Li, R.; Ma, T.; Leng, K.; Chen, Z.; Abdelwahab, I.; Loh, K. P. Exploring Ferroelectric Switching in α-$In_2Se_3$ for Neuromorphic Computing. *Advanced Functional Materials* **2020**, *30* (45). DOI: 10.1002/adfm.202004609.
(144) Liu, K.; Dang, B.; Zhang, T.; Yang, Z.; Bao, L.; Xu, L.; Cheng, C.; Huang, R.; Yang, Y. Multilayer Reservoir Computing Based on Ferroelectric

α-$In_2Se_3$ for Hierarchical Information Processing. *Advanced Materials* **2022**, *34* (48). DOI: 10.1002/adma.202108826.
(145) Tang, B.; Li, X.; Liao, J.; Chen, Q. Ultralow Power Consumption and Large Dynamic Range Synaptic Transistor Based on α-In2Se3 Nanosheets. *ACS Applied Electronic Materials* **2022**, *4* (2), 598-605, research-article. DOI: 10.1021/acsaelm.1c00970.
(146) Xiang, H.; Chien, Y.-C.; Li, L.; Zheng, H.; Li, S.; Duong, N. T.; Shi, Y.; Ang, K.-W. Enhancing Memory Window Efficiency of Ferroelectric Transistor for Neuromorphic Computing via Two-Dimensional Materials Integration. *Advanced Functional Materials* **2023**, *33* (42). DOI: 10.1002/adfm.202304657.
(147) Dutta, S.; Ye, H.; Khandker, A. A.; Kirtania, S. G.; Khanna, A.; Ni, K. Logic Compatible High-Performance Ferroelectric Transistor Memory. *IEEE Electron Device Letters* **2022**, *43* (3), 382-385. DOI: 10.1109/LED.2022.3148669.
(148) Tsai, S.-H.; Fang, Z.; Wang, X.; Chand, U.; Chen, C.-K.; Hooda, S.; Sivan, M.; Pan, J.; Zamburg, E.; Thean, A. V.-Y. Stress-Memorized HZO for High-Performance Ferroelectric Field-Effect Memtransistor. *ACS Applied Electronic Materials* **2022**, *4* (4), 1642-1650, research-article. DOI: 10.1021/acsaelm.1c01321.
(149) Hoffman, J.; Pan, X.; Reiner, J. W.; Walker, F. J.; Han, J. P.; Ahn, C. H.; Ma, T. P. Ferroelectric Field Effect Transistors for Memory Applications. *Advanced Materials* **2010**, *22* (26-27), 2957-2961. DOI: https://doi.org/10.1002/adma.200904327.
(150) Ma, T. P.; Jin-Ping, H. Why is nonvolatile ferroelectric memory field-effect transistor still elusive? *IEEE Electron Device Letters* **2002**, *23* (7), 386-388. DOI: 10.1109/LED.2002.1015207.
(151) Baek, S.; Yoo, H. H.; Ju, J. H.; Sriboriboon, P.; Singh, P.; Niu, J.; Park, J.-H.; Shin, C.; Kim, Y.; Lee, S. Ferroelectric Field-Effect-Transistor Integrated with Ferroelectrics Heterostructure. *Advanced Science* **2022**, *9* (21), 2200566. DOI: https://doi.org/10.1002/advs.202200566.
(152) Bian, J.; Cao, Z.; Zhou, P. Neuromorphic computing: Devices, hardware, and system application facilitated by two-dimensional materials. *Applied Physics Reviews* **2021**, *8* (4). DOI: 10.1063/5.0067352.
(153) Derbenwick, G.; Brewer, J. Alternative Memory Technologies. In *Nonvolatile Memory Technologies with Emphasis on Flash: A Comprehensive Guide to*

*Understanding and Using NVM Devices*, Wiley IEEE-Press, 2008; pp 617-740.
(154) Aziz, A.; Thirumala, S. K.; Wang, D.; George, S.; Li, X.; Datta, S.; Narayanan, V.; Gupta, S. K. Sensing in Ferroelectric Memories and Flip-Flops. In *Sensing of Non-Volatile Memory Demystified*, Ghosh, S. Ed.; Springer, 2018; pp 47-80.

(155) Deng, S.; Zhao, Z.; Kurinec, S. Overview of Ferroelectric Memory Devices and Reliability Aware Design Optimization. In *Proceedings of the 2021 on Great Lakes Symposium on VLSI*, 2021. DOI: 10.1145/3453688.3461743.
(156) Wang, D.; George, S.; Aziz, A.; Datta, S.; Narayanan, V.; Gupta, S. K. Ferroelectric Transistor based Non-Volatile Flip-Flop. In *International Symposium on Low Power Electronics and Design*, 2016; ISLPED. DOI: 10.1145/2934583.2934603.
(157) Xia, X.; Huang, W.; Hang, P.; Guo, T.; Yan, Y.; Yang, J.; Yang, D.; Yu, X.; Li, X. a. 2D-Material-Based Volatile and Nonvolatile Memristive Devices for Neuromorphic Computing. *ACS Materials Lett* **2023**, *5* (4), 1109-1135, review-article. DOI: 10.1021/acsmaterialslett.2c01026.
(158) Covi, E.; Mulaosmanovic, H.; Max, B.; Slesazeck, S.; Mikolajick, T. Ferroelectric-based synapses and neurons for neuromorphic computing - IOPscience. *Neuromorphic Computing and Engineering* **2022**, *2* (1), Text. DOI: 10.1088/2634-4386/ac4918.
(159) Hsiang, K.-Y.; Liao, C.-Y.; Chen, K.-T.; Lin, Y.-Y.; Chueh, C.-Y.; Chang, C.; Tseng, Y.-J.; Yang, Y.-J.; Chang, S. T.; Liao, M.-H.; et al. Ferroelectric HfZrO2 With Electrode Engineering and Stimulation Schemes as Symmetric Analog Synaptic Weight Element for Deep Neural Network Training. *IEEE Transactions on Electron Devices* **2020**, *67* (10), 4201-4207. DOI: 10.1109/TED.2020.3017463.
(160) Fu, X.; Li, T.; Cai, B.; Miao, J.; Panin, G. N.; Ma, X.; Wang, J.; Jiang, X.; Li, Q.; Dong, Y.; et al. Graphene/MoS2−xOx/graphene photomemristor with tunable non-volatile responsivities for neuromorphic vision processing. *Light:Science and Applications* **2023**, *12* (1). DOI: 10.1038/s41377-023-01079-5.
(161) Liu, X.; Katti, K.; Jariwala, D. Accelerate and actualize: Can 2D materials bridge the gap between neuromorphic hardware and the human brain? *Matter* **2023**, *6* (5), 1348-1365. DOI: 10.1016/j.matt.2023.03.016.
(162) Haizhong Guo and Meng He and Can Wang and Lin Gu and Xiulai Xu and Guangyu Zhang and Guozhen Yang and Kuijuan Jin and Chen, Ge, J. D. a. D. X. a. Q. Z. a. H. Z. a. F. M. a. X. F. a. Q. S. a. H. N. a. T. L. a. E.-j. G. a. A robust neuromorphic vision sensor with optical control of ferroelectric switching. *Nano Energy* **2021**, *89*, 106439. DOI: https://doi.org/10.1016/j.nanoen.2021.106439.
(163) Xue, F.; He, X.; Liu, W.; Periyanagounder, D.; Zhang, C.; Chen, M.; Lin, C.-H.; Luo, L.; Yengel, E.; Tung, V.; et al. Optoelectronic Ferroelectric Domain-Wall Memories Made from a Single Van Der Waals Ferroelectric. *Advanced Functional Materials* **2020**, *30* (52), 2004206. DOI: https://doi.org/10.1002/adfm.202004206.
(164) Kittel, C. *Introduction to Solid State Physics*; John Wiley & Sons, 1976.
(165) Endoh, T.; Koike, H.; Ikeda, S.; Hanyu, T.; Ohno, H. An Overview of Nonvolatile Emerging Memories— Spintronics for Working Memories. *IEEE Journal on Emerging and Selected Topics in Circuits and Systems* **2016**, *6* (2), 109-119. DOI: 10.1109/JETCAS.2016.2547704.
(166) Su, Y.; Li, X.; Zhu, M.; Zhang, J.; You, L.; Tsymbal, E. Y. Van der Waals Multiferroic Tunnel Junctions. *Nano Letters* **2020**, *21* (1), 175-181, rapid-communication. DOI: 10.1021/acs.nanolett.0c03452.
(167) Lukashev, P. V.; Burton, J. D.; Jaswal, S. S.; Tsymbal, E. Y. Ferroelectric control of the

magnetocrystalline anisotropy of the Fe/BaTiO3(001) interface. *Journal of Physics: Condensed Matter* **2012**, *24* (22), Text. DOI: 10.1088/0953-8984/24/22/226003.

(168) Yu, C.; Cheng, J.; Zhang, Y.; Liu, Z.; Liu, X.; Jia, C.; Li, X.; Yang, J. Two-Dimensional Os2Se3 Nanosheet: A Ferroelectric Metal with Room-Temperature Ferromagnetism. *Journal of Physics: Chemistry Letters* **2024**, *15*, rapid-communication. DOI: 10.1021/acs.jpclett.4c00524.

(169) Lai, Y.; Song, Z.; Yi Wan a, M. X. a., Changsheng Wang; Ye, u.; Dai, L.; Zhang, Z.; Yang, W.; Du, H.; Yang, J. Two-dimensional ferromagnetism and driven ferroelectricity in van der Waals CuCrP2S6. *Nanoscale* **2019**, *11*, 5163-5170. DOI: 10.1039/C9NR00738E.

(170) Wang, X.; Shang, Z.; Zhang, C.; Kang, J.; Liu, T.; Wang, X.; Chen, S.; Liu, H.; Tang, W.; Zeng, Y.-J.; et al. Electrical and magnetic anisotropies in van der Waals multiferroic CuCrP2S6. *Nature Communications* **2023**, *14* (1), 1-8, OriginalPaper. DOI: 10.1038/s41467-023-36512-1.

(171) Zheng, S.; Wu, E.; Feng, Z.; Rao Zhang a, Y. X. a., Yuanyuan Yu a, Rui Zhang; Li, Q.; Liu, J.; Pang, W.; Zhang, H.; Zhang, D. Acoustically enhanced photodetection by a black phosphorus–MoS2 van der Waals heterojunction p–n diode. *Nanoscale* **2018**, *10*. DOI: 10.1039/C8NR02022A.

(172) Datta, K.; Li, Z.; Lyu, Z.; Deotare, P. B. Piezoelectric Modulation of Excitonic Properties in Monolayer WSe2 under Strong Dielectric Screening. *ACS Nano* **2021**, *15* (7), 12334–12341, research-article. DOI: 10.1021/acsnano.1c04269.

(173) Du, X.; Tang, Z.; Leblanc, C.; Jariwala, D.; Olsson, R. H. High-Performance SAW Resonators at 3 GHz Using AlScN on a 4H-SiC Substrate. In Joint Conference of the European Frequency and Time Forum and IEEE International Frequency Control Symposium, 2022.

(174) Preciado, E.; Schülein, F. J. R.; Nguyen, A. E.; Barroso, D.; Isarraraz, M.; von Son, G.; Lu, I.-H.; Michailow, W.; Möller, B.; Klee, V.; et al. Scalable fabrication of a hybrid field-effect and acousto-electric device by direct growth of monolayer MoS2/LiNbO3. *Nature Communications* **2015**, *6* (1), OriginalPaper. DOI: 10.1038/ncomms9593.

(175) Guan, H.; Hong, J.; Wang, X.; Ming, J.; Zhang, Z.; Liang, A.; Han, X.; Dong, J.; Qiu, W.; Chen, Z.; et al. Broadband, High-Sensitivity Graphene Photodetector Based on Ferroelectric Polarization of Lithium Niobate. *Advanced Optical Materials* **2021**, *9* (16), 2100245. DOI: https://doi.org/10.1002/adom.202100245.

(176) Ke, C.; Huang, J.; Liu, S. Two-dimensional ferroelectric metal for electrocatalysis. *Materials Horizons* **2021**, *8* (12), 3387-3393. DOI: 10.1039/D1MH01556G.

(177) Song, S.; Jeon, S.; Rahaman, M.; Lynch, J.; Rhee, D.; Kumar, P.; Chakravarthi, S.; Kim, G.; Du, X.; Blanton, E. W.; et al. Wafer-scale growth of two-dimensional, phase-pure InSe. *Matter* **2023**, *6* (10), 3483-3498. DOI: https://doi.org/10.1016/j.matt.2023.07.012.

(178) Wei, T.-R.; Jin, M.; Wang, Y.; Chen, H.; Gao, Z.; Zhao, K.; Qiu, P.; Shan, Z.; Jiang, J.; Li, R.; et al. Exceptional plasticity in the bulk single-crystalline van der Waals semiconductor InSe. *Science* **2020**, *369* (6503), 542-545. DOI: doi:10.1126/science.aba9778.

(179) Wong, L. W.; Yang, K.; Han, W.; Zheng, X.; Wong, H. Y.; Tsang, C. S.; Lee, C.-S.; Lau, S. P.; Ly, T. H.; Yang, M.; et al. Deciphering the ultra-high plasticity in metal monochalcogenides. *Nature*

*Materials* **2024**. DOI: 10.1038/s41563-023-01788-7.
(180) Ge, B.; Li, C.; Lu, W.; Ye, H.; Li, R.; He, W.; Wei, Z.; Shi, Z.; Kim, D.; Zhou, C.; et al. Dynamic Phase Transition Leading to Extraordinary Plastic Deformability of Thermoelectric SnSe2 Single Crystal. *Advanced Energy Materials* **2023**, *13* (27), 2300965. DOI: https://doi.org/10.1002/aenm.202300965.
(181) Jin, G.; Lee, C.-S.; Liao, X.; Kim, J.; Wang, Z.; Okello, O. F. N.; Park, B.; Park, J.; Han, C.; Heo, H.; et al. Atomically thin three-dimensional membranes of van der Waals semiconductors by wafer-scale growth. *Science Advances* **2019**, *5* (7), eaaw3180. DOI: doi:10.1126/sciadv.aaw3180.
(182) Yan, S.; Xu, C.; Zhong, C.; Chen, Y.; Che, X.; Luo, X.; Zhu, Y. Phase Instability in van der Waals In2Se3 Determined by Surface Coordination. *Angewandte Chemie International Edition* **2023**, *62* (17), e202300302. DOI: https://doi.org/10.1002/anie.202300302.
(183) Ignacio, N. D.; Fatheema, J.; Jeon, Y.-R.; Akinwande, D. Air-Stable Atomically Encapsulated Crystalline-Crystalline Phase Transitions in In2Se3. *Advanced Electronic Materials* **2024**, *10* (1), 2300457. DOI: https://doi.org/10.1002/aelm.202300457.
(184) Ye, F.; Lee, J.; Hu, J.; Mao, Z.; Wei, J.; Feng, P. X.-L. Environmental Instability and Degradation of Single- and Few-Layer WTe2 Nanosheets in Ambient Conditions. *Small* **2016**, *12* (42), 5802-5808. DOI: https://doi.org/10.1002/smll.201601207.
(185) Song, S.; Kim, S.-Y.; Kwak, J.; Jo, Y.; Kim, J. H.; Lee, J. H.; Lee, J.-U.; Kim, J. U.; Yun, H. D.; Sim, Y.; et al. Electrically Robust Single-Crystalline WTe2 Nanobelts for Nanoscale Electrical Interconnects. *Advanced Science* **2019**, *6* (3), 1801370. DOI: https://doi.org/10.1002/advs.201801370.
(186) Xue, W.; Jiang, Q.; Wang, F.; He, R.; Pang, R.; Yang, H.; Wang, P.; Yang, R.; Zhong, Z.; Zhai, T.; et al. Discovery of Robust Ferroelectricity in 2D Defective Semiconductor α-Ga2Se3. *Small* **2022**, *18* (8), 2105599. DOI: https://doi.org/10.1002/smll.202105599.
(187) Jiang, A. Q.; Liu, X. B.; Zhang, Q. Nanosecond-range imprint and retention characterized from polarization-voltage hysteresis loops in insulating or leaky ferroelectric thin films. *Applied Physics Letters* **2011**, *99* (14). DOI: 10.1063/1.3647577 (acccessed 1/31/2024).
(188) Martin, L. W.; Maria, J.-P.; Schlom, D. G. Lifting the fog in ferroelectric thin-film synthesis. *Nature Materials* **2024**, *23* (1), 9-10. DOI: 10.1038/s41563-023-01732-9.
(189) Chin, H.-T.; Klimes, J.; Hu, I. F.; Chen, D.-R.; Nguyen, H.-T.; Chen, T.-W.; Ma, S.-W.; Hofmann, M.; Liang, C.-T.; Hsieh, Y.-P. Ferroelectric 2D ice under graphene confinement. *Nature Communications* **2021**, *12* (1), 6291. DOI: 10.1038/s41467-021-26589-x.
(190) Lanza, M.; Waser, R.; Ielmini, D.; Yang, J. J.; Goux, L.; Suñe, J.; Kenyon, A. J.; Mehonic, A.; Spiga, S.; Rana, V.; et al. Standards for the Characterization of Endurance in Resistive Switching Devices. *ACS Nano* **2021**, *15* (11), 17214–17231, review-article. DOI: 10.1021/acsnano.1c06980.
(191) Okuno, J.; Yonai, T.; Kunihiro, T.; Shuto, Y.; Alcala, R.; Lederer, M.; Seidel, K.; Mikolajick, T.; Schroeder, U.; Tsukamoto, M.; et al. Investigation of Recovery Phenomena in Hf0.5Zr0.5O2-Based 1T1C FeRAM. *IEEE Journal of the Electron Devices Society* **2022**, *11*, 43-46. DOI: 10.1109/JEDS.2022.3230402.
(192) Baksa, J. C. S. M.; Behrendt, D.; Calderon, S.; Goodling, D.; Hayden, J.; He, F.; Jacques, L.; Lee, S.

H.; Smith, W.; Suceava, A.; et al. Perspectives and progress on wurtzite ferroelectrics: Synthesis, characterization, theory, and device applications. *Applied Physics Letters* **2024**, *124* (8). DOI: 10.1063/5.0185066.
(193) Wu, P.; Zhang, T.; Zhu, J.; Palacios, T.; Kong, J. 2D materials for logic device scaling. *Nature Materials* **2024**, *23* (1), 23-25, BriefCommunication. DOI: 10.1038/s41563-023-01715-w.
(194) Li, T.; Sharma, P.; Lipatov, A.; Lee, H.; Lee, J.-W.; Zhuravlev, M. Y.; Paudel, T. R.; Genenko, Y. A.; Eom, C.-B.; Tsymbal, E. Y.; et al. Polarization-Mediated Modulation of Electronic and Transport Properties of Hybrid MoS2–BaTiO3–SrRuO3 Tunnel Junctions. *Nano Letters* **2017**, *17* (2), 922-927, rapid-communication. DOI: 10.1021/acs.nanolett.6b04247.
(195) KC, S.; Longo, R. C.; Wallace, R. M.; Cho, K. Computational Study of MoS2/HfO2 Defective Interfaces for Nanometer-Scale Electronics. *ACS Omega* **2017**, *2* (6), 2827–2834, research-article. DOI: 10.1021/acsomega.7b00636.
(196) Kohlstedt, H.; Mustafa, Y.; Gerber, A.; Petraru, A.; Fitsilis, M.; Meyer, R.; Böttger, U.; Waser, R. Current status and challenges of ferroelectric memory devices Author links open overlay panel. *Microelectronic Engineering* **2005**, *80*, 296-304. DOI: 10.1016/j.mee.2005.04.084.